\documentclass[11pt,a4paper]{article}

\usepackage[english]{babel}

\usepackage[a4paper,margin=2.5cm]{geometry}

\usepackage{amsmath,amssymb,amsthm,mathtools}

\usepackage{graphicx,booktabs,subcaption,float,xcolor,csquotes}

\usepackage{setspace}

\usepackage{enumitem}

\usepackage{titling}
\usepackage[
  backend=biber,
  style=apa,
  doi=true,
  url=true,
  bibencoding=utf8
]{biblatex}
\usepackage[
  colorlinks=true,
  linkcolor=blue,
  citecolor=blue,
  urlcolor=blue,
  linktocpage=true,
  hypertexnames=false
]{hyperref}
\usepackage{xurl}

\usepackage{proof-at-the-end}

\theoremstyle{definition}
\newtheorem{definition}{Definition}[section]

\newtheorem{assumption}{Assumption}
\theoremstyle{remark}

\newif\ifproofsinappendix
\proofsinappendixtrue
\ifproofsinappendix
  \pratendSetGlobal{no restate, proof at the end, link to proof,
    text link={\noindent See \hyperref[proof:prAtEnd\pratendcountercurrent]{proof} on page~\pageref{proof:prAtEnd\pratendcountercurrent}.}}
\else
  \pratendSetGlobal{normal}
\fi

\title{Partial Identification of Survival Prediction Performance under Dependent Right Censoring}

\author{%
  Yuji Komiyama\thanks{Email: \texttt{komiyama.yuji.r2@dc.tohoku.ac.jp}}
  \qquad
  Yasumasa Matsuda\thanks{Corresponding author. Email: \texttt{yasumasa.matsuda.a4@tohoku.ac.jp}}
  \\[0.4em]
  {\normalsize Graduate School of Economics and Management, Tohoku University, Japan}
}

\date{}

\begin{document}

\maketitle

\begin{abstract}
  Survival prediction models are usually evaluated under right censoring by inverse probability of censoring weighting (IPCW), which assumes independent censoring, an assumption that cannot generally be verified from right-censored data.
  Leaving the dependence between event and censoring times unrestricted, we derive sharp bounds on the integrated Brier score (IBS) of a fixed model and on IBS differences between models by partial identification.
  The bounds use the fact that each individual's event time is shared across evaluation times and models, and for more than two models we also derive the sharp upper bound on the difference from the best competitor.
  A negative upper bound on a difference shows that one model has a smaller IBS than the other under every full-data distribution compatible with the observed data.
  For a fixed observed-data distribution, we develop estimators, confidence regions, a test of robust strict dominance, and an intersection--union test.
  In simulations, the confidence regions and pairwise tests performed close to their nominal levels.
  In three applications, the tests supported the IPCW ranking only for mortgage prepayment data with little censoring, whereas for the other two data sets the observed data alone did not support it.
\end{abstract}


\section{Introduction}\label{sec:introduction}

Evaluating a survival prediction model requires each individual's survival status at each time, but right censoring leaves part of it unobserved.
Such a model uses individual characteristics, called covariates, to predict the probability that an event of interest, such as death or disease recurrence, has not occurred by each time.
Its performance is often measured by the Brier score at a given time, the expected squared difference between this predicted probability and the indicator that the event has not occurred by that time \parencite{1999Graf_AssessmentComparisonPrognostica,2006Gerds_ConsistentEstimationExpected}.
The integrated Brier score (IBS) averages the Brier score over a prespecified evaluation period.
These measures are used to compare candidate models, such as the Cox proportional hazards model \parencite{1972Cox_RegressionModelsLifeTables}, random survival forests \parencite{2008Ishwaran_RandomSurvivalForests}, and DeepSurv \parencite{2018Katzman_DeepSurvPersonalized}, and to choose the one with the smallest IBS.
When follow-up ends before the event, however, the observation is right censored: the event time is known only to exceed the censoring time, so the survival status after censoring is unknown.
Evaluating the Brier score and the IBS therefore requires accounting for the unobserved event times.

Previous studies have estimated prediction performance from right-censored data by inverse probability of censoring weighting (IPCW) under independent censoring.
IPCW weights each individual whose survival status is observed by the inverse of the probability of remaining uncensored until that status is observed.
For weights based on the marginal censoring distribution, independent censoring means that the censoring time is independent of both the event time and the covariates.
\textcite{1999Graf_AssessmentComparisonPrognostica} proposed such an estimator of the Brier score, and \textcite{2006Gerds_ConsistentEstimationExpected} proved it consistent and extended it to conditionally independent censoring, under which the two times are independent given the covariates.
Even this condition fails when an unobserved factor, such as a deterioration in health, makes a patient both more likely to drop out and more likely to die soon.
IPCW estimates of the IBS are then biased, and the ranking of models based on them can differ from the ranking under the full-data distribution.
Moreover, neither assumption can generally be verified from right-censored data, because different dependence structures can generate the same observed-data distribution \parencite{1975Tsiatis_NonidentifiabilityAspectProblem}.

Few studies have addressed the evaluation of survival predictions under dependent censoring.
\textcite{2026Lillelund_OvercomingDependentCensoring} propose a dependent Brier score, which they describe as the first such measure.
Under an assumed Archimedean copula, which specifies the dependence between event and censoring times, they replace the unknown event time of each censored individual with its conditional mean beyond the censoring time.
Their score therefore targets a surrogate defined by these imputed times rather than the Brier score defined by the true event times.
It also requires specifying the copula family and its parameter, although the dependence structure is generally not identifiable from right-censored data.
A method that evaluates the IBS based on the true event times without specifying this dependence is therefore needed.

In this paper, we evaluate survival predictions by partial identification, leaving the dependence between event and censoring times unrestricted.
Partial identification characterizes the set of values of a target quantity that are compatible with the observed data and the maintained conditions \parencite{2003Manski_PartialIdentificationProbability,2010Tamer_PartialIdentificationEconometrics,2020Molinari_MicroeconometricsPartialIdentification}.
Under dependent censoring, it has yielded bounds on survival functions \parencite{1976Peterson_BoundsJointDistribution} and on regression parameters and covariate effects \parencite{2024Sakaguchi_PartialIdentificationInferencea,2026Willems_BoundsRegressionParametersa}, but not on the prediction performance of a given model.
In our setting, many full-data distributions of the event time, censoring time, and covariates generate the same observed-data distribution, yet they can imply different values of the IBS, our measure of prediction performance.
The set of these values is called the identified set, and its bounds describe the range of performance supported by the observed data.
The identified set thus shows what can be concluded about prediction performance without assuming independent censoring or specifying a copula.

Partial identification has also been used to evaluate prediction performance when outcomes are only partly observed.
\textcite{2026Rambachan_RobustDesignEvaluation} bound performance measures under selective observation of outcomes, allowing unobserved confounding subject to restrictions on outcome differences between selected and unselected individuals with the same covariates.
\textcite{2024Polo_WeakSupervisionPerformance} bound the accuracy and related measures of a fixed classifier using observed weak labels and a model linking them to the unobserved true labels.
\textcite{2024Guerdan_PredictivePerformanceComparison} bound the difference in performance between two decision policies directly, using the fact that the policies share the same unobserved outcomes, which avoids counting the uncertainty common to both twice.
\textcite{2026Liu_IdentificationInferenceAlgorithmic} characterize the sharp identification region of the fairness--accuracy frontier under selective labels for binary outcomes, measuring accuracy by classification error and fairness by differences in positive decision rates between groups.
These studies do not explicitly address right-censored survival outcomes, for which a single event time determines survival status across all evaluation times.
We derive sharp bounds on the IBS and on differences between models while respecting this common event time across evaluation times and models.

Our main contribution is to derive sharp bounds on the IBS and on IBS differences between models, using the fact that each individual's event time is shared across evaluation times and models.
Sharpness means that the bounds cannot be narrowed using the observed-data distribution alone.
Bounding the Brier score separately at each time would allow survival statuses that no single event time produces, and bounding each model separately can widen the bounds on an IBS difference.
When more than two models are compared, we also derive the sharp upper bound on the difference between each model's IBS and that of its best competitor.
For models trained separately from the evaluation sample and a fixed observed-data distribution, we then develop estimators of these bounds, confidence regions, a test of robust strict dominance, and an intersection--union test.
A negative upper bound on one model's IBS minus another's establishes a smaller IBS for the first model under every compatible full-data distribution.
If the two bounds have opposite signs, the observed data alone cannot rank the two models.

The remainder of this paper is organized as follows.
Section~\ref{sec:partial-identification} derives the bounds for a single model, two models, and more than two models, and Section~\ref{sec:estimation-inference} develops estimators, confidence regions, and tests.
Section~\ref{sec:simulation} examines their finite-sample behavior through simulations, and Section~\ref{sec:real-data} applies them to three real data sets.
Section~\ref{sec:conclusion} concludes, and Appendices~\ref{app:proofs-partial-identification} and~\ref{app:proofs-inference} contain the proofs.

\section{Partial identification of prediction performance}\label{sec:partial-identification}
\pratendSetLocal{category=partial-identification}

This section derives sharp bounds on prediction performance without restricting the dependence between event and censoring times.
Section~\ref{subsec:framework} introduces the setting, the evaluation targets, and the identified set.
Sections~\ref{subsec:single-model} and~\ref{subsec:two-models} derive sharp lower and upper bounds on the IBS of a single model and on the IBS difference between two models.
Section~\ref{subsec:multiple-models} derives the sharp upper bound on the difference from the best competitor and uses it to define when a model is robustly strictly best among the $K$ models.

\subsection{Framework and evaluation targets}\label{subsec:framework}

For each individual, we consider the event time $T\ge 0$, the censoring time $C\ge 0$, and the covariates $X\in\mathcal X$, but we observe only the observed time, the event indicator, and the covariates:
\[
    Y=\min(T,C),
    \qquad
    \Delta=\mathbf 1\{T\le C\},
    \qquad
    X.
\]
We write the observed data for one individual as $O=(Y,\Delta,X)$ and the evaluation sample as $O_1,\ldots,O_n$.
If $\Delta=1$, the event occurs during follow-up, and its time is observed as $T=Y$.
If $\Delta=0$, the individual is censored at $C=Y$, and the event time satisfies $T>Y$ but is otherwise unknown.
Hence, for a censored individual, the data reveal only that the event time is later than the observed time.

We distinguish two distributions: the full-data distribution $Q$ of $(T,C,X)$ and the observed-data distribution $P$ of $(Y,\Delta,X)$.
Each full-data distribution determines an observed-data distribution.
We say that $Q$ is compatible with $P$ if the distribution of $(\min(T,C),\mathbf 1\{T\le C\},X)$ under $Q$ equals $P$, and we denote by $\mathcal Q(P)$ the set of full-data distributions compatible with $P$.
We impose no restriction on the dependence between $T$ and $C$, so a single observed-data distribution is generally compatible with multiple full-data distributions.

Let $0<\tau<\infty$ be a prespecified evaluation horizon, defining the evaluation period $[0,\tau]$.
We measure prediction performance by the Brier score at each time $t$ and by the IBS, its average over this period.
A prediction model outputs a predicted survival probability $\widehat S(t\mid x)\in[0,1]$ for each time $t$ and covariate value $x$, measurable in $(t,x)$.
Like a survival function, $\widehat S(t\mid x)$ is nonincreasing in $t$ for every $x$.
The model is trained on data separate from the evaluation sample and is held fixed during evaluation.
Under a full-data distribution $Q$, the Brier score at time $t$ and the IBS are
\begin{equation}
    \operatorname{BS}(t;Q)
    =
    \mathbb E_Q\Bigl[\bigl\{\mathbf 1(T>t)-\widehat S(t\mid X)\bigr\}^2\Bigr],
    \qquad
    \operatorname{IBS}(\tau;Q)
    =
    \frac{1}{\tau}\int_0^\tau\operatorname{BS}(t;Q)\,dt.
    \label{eq:bs-ibs}
\end{equation}

A single event time determines the survival status at every evaluation time: $\mathbf 1(T>t)$ equals one before the event time and zero afterward.
To make this explicit, write $r=\min(T,\tau)$ for an individual's event time truncated at the evaluation horizon.
Define the loss for truncated event time $r$ and covariates $x$ by
\begin{equation}
    L(r,x;\tau)
    =
    \int_0^r\bigl\{1-\widehat S(t\mid x)\bigr\}^2\,dt
    +\int_r^\tau\widehat S(t\mid x)^2\,dt,
    \qquad
    0\le r\le\tau.
    \label{eq:individual-loss}
\end{equation}
When several models are compared, we index them by $j$ and write $\widehat S_j$, $\operatorname{BS}_j$, $\operatorname{IBS}_j$, and $L_j$.
Exchanging the expectation and the integral in~\eqref{eq:bs-ibs} gives $\operatorname{IBS}(\tau;Q)=\tau^{-1}\mathbb E_Q L(\min(T,\tau),X;\tau)$.

The Brier scores and IBS values depend on the full-data distribution $Q$ and therefore are generally not uniquely determined by the observed-data distribution $P$ alone.
Under conditionally independent censoring, $T\perp C\mid X$, the Brier score is uniquely determined by $P$ given suitable positivity conditions.
In that case, IPCW can use the censoring distribution given $X$ \parencite{2006Gerds_ConsistentEstimationExpected}.

For a target quantity $\theta(Q)$, such as $\operatorname{IBS}(\tau;Q)$ or $\operatorname{IBS}_j(\tau;Q)-\operatorname{IBS}_k(\tau;Q)$, we define its identified set as the set of values that the target takes under the full-data distributions compatible with $P$:
\[
    \Theta_{\mathrm I}(P)
    =
    \bigl\{\theta(Q):Q\in\mathcal Q(P)\bigr\}.
\]
The observed data cannot distinguish among the values in $\Theta_{\mathrm I}(P)$.
A set $B(P)$ is an outer set if $\Theta_{\mathrm I}(P)\subseteq B(P)$.
Such a set contains $\theta(Q)$ for every $Q\in\mathcal Q(P)$, but it may also include values that cannot be attained under any compatible full-data distribution.
An outer set is said to be sharp if $B(P)=\Theta_{\mathrm I}(P)$.

We define the sharp lower and upper bounds on $\theta$ by
\[
    \underline\theta(P)=\inf_{Q\in\mathcal Q(P)}\theta(Q),
    \qquad
    \overline\theta(P)=\sup_{Q\in\mathcal Q(P)}\theta(Q).
\]
These are, respectively, the largest lower bound and the smallest upper bound that hold for every $Q\in\mathcal Q(P)$.

In the following subsections, we express the sharp lower and upper bounds in terms of $P$ for the IBS of a single model and for the IBS difference between two models.
For comparisons among $K\ge3$ models, we derive the sharp upper bound on the difference from the best competitor and use its sign to assess whether a model is robustly strictly best.

\subsection{Single model}\label{subsec:single-model}

At a fixed time \(t>0\), the observed data determine the survival status \(\mathbf 1(T>t)\) except for individuals censored before \(t\). Specifically, the survival status equals one if \(Y>t\) or \((Y=t,\Delta=0)\), and zero if \(Y\le t\) and \(\Delta=1\). For individuals with \(Y<t\) and \(\Delta=0\), both survival states are compatible with the observed data, so the squared prediction loss can be either \(\widehat S(t\mid X)^2\) or \(\{1-\widehat S(t\mid X)\}^2\). Taking the minimum and maximum of these losses yields the following bounds, which are sharp when \(P(Y=t,\Delta=0)=0\).
\begin{theoremEnd}{proposition}\label{prop:bs-bounds}
    Fix $t>0$ and define
    \begin{align*}
        \underline{\operatorname{BS}}(t;P)
        =
        \mathbb E_P\Bigl[
         & \mathbf 1(Y>t)\bigl\{1-\widehat S(t\mid X)\bigr\}^2
        +\mathbf 1(Y\le t,\Delta=1)\,\widehat S(t\mid X)^2                                                          \\
         & +\mathbf 1(Y\le t,\Delta=0)\min\bigl\{\widehat S(t\mid X)^2,\bigl\{1-\widehat S(t\mid X)\bigr\}^2\bigr\}
            \Bigr],
    \end{align*}
    \begin{align*}
        \overline{\operatorname{BS}}(t;P)
        =
        \mathbb E_P\Bigl[
         & \mathbf 1(Y>t)\bigl\{1-\widehat S(t\mid X)\bigr\}^2
        +\mathbf 1(Y\le t,\Delta=1)\,\widehat S(t\mid X)^2                                                          \\
         & +\mathbf 1(Y\le t,\Delta=0)\max\bigl\{\widehat S(t\mid X)^2,\bigl\{1-\widehat S(t\mid X)\bigr\}^2\bigr\}
            \Bigr].
    \end{align*}
    Then the following hold.
    \begin{enumerate}[label=(\roman*)]
        \item \emph{Outer set.}
              The interval $[\underline{\operatorname{BS}}(t;P),\overline{\operatorname{BS}}(t;P)]$ is an outer set for the identified set of $\operatorname{BS}(t;Q)$; that is,
              \[
                  \underline{\operatorname{BS}}(t;P)\le\operatorname{BS}(t;Q)\le\overline{\operatorname{BS}}(t;P)
                  \quad\text{for every }Q\in\mathcal Q(P).
              \]
        \item \emph{Sharpness.}
              If $P(Y=t,\Delta=0)=0$, then $\underline{\operatorname{BS}}(t;P)$ and $\overline{\operatorname{BS}}(t;P)$ are the sharp lower and upper bounds on $\operatorname{BS}(t;Q)$, and both are attained.
              That is, the identified set of $\operatorname{BS}(t;Q)$ is the closed interval $[\underline{\operatorname{BS}}(t;P),\overline{\operatorname{BS}}(t;P)]$.
        \item \emph{Width.}
              The width of the bounds is
              \[
                  \overline{\operatorname{BS}}(t;P)-\underline{\operatorname{BS}}(t;P)
                  =
                  \mathbb E_P\bigl[\mathbf 1(Y\le t,\Delta=0)\,\bigl|1-2\widehat S(t\mid X)\bigr|\bigr].
              \]
    \end{enumerate}
\end{theoremEnd}
\begin{proofEnd}
    \emph{Part (i).}
    Let $Q\in\mathcal Q(P)$.
    By the definition of compatibility in Section~\ref{subsec:framework}, the observed data $O=(Y,\Delta,X)$ have distribution $P$ under $Q$.
    Write $\underline b_{\mathrm{BS}}(t,O)$ and $\overline b_{\mathrm{BS}}(t,O)$ for the expressions inside the expectations defining $\underline{\operatorname{BS}}(t;P)$ and $\overline{\operatorname{BS}}(t;P)$, respectively.

    If $Y>t$, then $T\ge Y>t$, so the loss equals $\{1-\widehat S(t\mid X)\}^2$.
    If $Y\le t$ and $\Delta=1$, then $T=Y\le t$, so the loss equals $\widehat S(t\mid X)^2$.
    In these two cases, both $\underline b_{\mathrm{BS}}(t,O)$ and $\overline b_{\mathrm{BS}}(t,O)$ equal the loss.
    If $Y\le t$ and $\Delta=0$, then $\mathbf 1(T>t)\in\{0,1\}$, so the loss $\{\mathbf 1(T>t)-\widehat S(t\mid X)\}^2$ lies between the minimum and the maximum of $\widehat S(t\mid X)^2$ and $\{1-\widehat S(t\mid X)\}^2$.
    Thus, in all three cases,
    \[
        \underline b_{\mathrm{BS}}(t,O)
        \le \{\mathbf 1(T>t)-\widehat S(t\mid X)\}^2
        \le \overline b_{\mathrm{BS}}(t,O)
        \qquad Q\text{-almost surely}.
    \]

    Since $\underline b_{\mathrm{BS}}(t,O)$ and $\overline b_{\mathrm{BS}}(t,O)$ are functions of $O$ alone, compatibility implies that their expectations under $Q$ equal their expectations under $P$.
    Taking expectations of the loss inequalities therefore gives
    \begin{align*}
        \underline{\operatorname{BS}}(t;P)
         & = \mathbb E_P[\underline b_{\mathrm{BS}}(t,O)]
        = \mathbb E_Q[\underline b_{\mathrm{BS}}(t,O)]    \\
         & \le \operatorname{BS}(t;Q)
        \le \mathbb E_Q[\overline b_{\mathrm{BS}}(t,O)]   \\
         & = \mathbb E_P[\overline b_{\mathrm{BS}}(t,O)]
        = \overline{\operatorname{BS}}(t;P).
    \end{align*}

    \emph{Part (ii).}
    Suppose that $P(Y=t,\Delta=0)=0$.
    We first construct a compatible full-data distribution by choosing event times to minimize the squared prediction loss.
    For individuals censored before $t$, the loss is $\widehat S(t\mid X)^2$ if $T\le t$ and $\{1-\widehat S(t\mid X)\}^2$ if $T>t$.
    Since $\widehat S(t\mid X)^2\le\{1-\widehat S(t\mid X)\}^2$ if and only if $\widehat S(t\mid X)\le 1/2$, we choose the event times accordingly.
    Given $O=(Y,\Delta,X)$, define $C=Y$ and
    \[
        T=
        \begin{cases}
            Y,   & \Delta=1,                                    \\
            Y+1, & \Delta=0,\ Y\ge t,                           \\
            t,   & \Delta=0,\ Y<t,\ \widehat S(t\mid X)\le 1/2, \\
            t+1, & \Delta=0,\ Y<t,\ \widehat S(t\mid X)>1/2,
        \end{cases}
    \]
    and let $\underline Q$ be the distribution of $(T,C,X)$ when $O$ follows $P$.
    This construction is measurable because $\widehat S(t\mid X)$ is measurable.
    In every case, $T\le C$ if and only if $\Delta=1$, and $\min(T,C)=Y$; hence $\underline Q\in\mathcal Q(P)$.

    We now verify that $\underline Q$ attains the lower bound.
    For individuals with $Y<t$ and $\Delta=0$, the construction sets $T=t$ when $\widehat S(t\mid X)\le1/2$ and $T=t+1$ otherwise.
    Their loss is therefore $\min\{\widehat S(t\mid X)^2,\{1-\widehat S(t\mid X)\}^2\}$.
    For individuals with $Y>t$ or $Y\le t$ and $\Delta=1$, the survival status is determined by the observed data, and the corresponding loss agrees with the definition of $\underline{\operatorname{BS}}(t;P)$.
    Since $P(Y=t,\Delta=0)=0$, taking expectations gives $\operatorname{BS}(t;\underline Q)=\underline{\operatorname{BS}}(t;P)$.

    The upper bound is attained by reversing the loss assignments for individuals censored before $t$.
    Interchanging the values $t$ and $t+1$ in the last two cases of the definition of $T$ gives $\overline Q\in\mathcal Q(P)$ under which each such individual has the larger loss.
    The same probability-zero case and expectation argument give $\operatorname{BS}(t;\overline Q)=\overline{\operatorname{BS}}(t;P)$.

    Every value between the two attained bounds can also be attained by mixing $\underline Q$ and $\overline Q$.
    For $\lambda\in[0,1]$, define $Q_\lambda=(1-\lambda)\underline Q+\lambda\overline Q$.
    For every measurable set $A$ of observed-data values, compatibility gives
    \begin{align*}
        Q_\lambda(O\in A)
         & = (1-\lambda)\underline Q(O\in A)+\lambda\overline Q(O\in A) \\
         & = (1-\lambda)P(A)+\lambda P(A)
        = P(A).
    \end{align*}
    Thus $Q_\lambda$ also induces $P$ and belongs to $\mathcal Q(P)$.
    Since the Brier score is the expectation of a fixed loss, its value under $Q_\lambda$ is $(1-\lambda)\underline{\operatorname{BS}}(t;P)+\lambda\overline{\operatorname{BS}}(t;P)$.
    By part (i), no value outside this interval is attainable, so the identified set equals $[\underline{\operatorname{BS}}(t;P),\overline{\operatorname{BS}}(t;P)]$.
    Its endpoints are therefore its infimum and supremum, the sharp bounds defined in Section~\ref{subsec:framework}.

    \emph{Part (iii).}
    The definitions of $\underline{\operatorname{BS}}(t;P)$ and $\overline{\operatorname{BS}}(t;P)$ differ only in the term for $Y\le t$ and $\Delta=0$.
    For real numbers $u,v$, $\max\{u,v\}-\min\{u,v\}=|u-v|$, and $\{1-\widehat S(t\mid X)\}^2-\widehat S(t\mid X)^2=1-2\widehat S(t\mid X)$.
    Subtracting the two definitions therefore gives the stated width.
\end{proofEnd}

The width of these bounds depends not only on the proportion of individuals censored by $t$ but also on how far their predicted survival probabilities are from $1/2$.
By Proposition~\ref{prop:bs-bounds}(iii), each individual censored by $t$ contributes $|1-2\widehat S(t\mid X)|$, the absolute difference between its two candidate losses, to the bound width.
An individual with $\widehat S(t\mid X)=1/2$ incurs the loss $1/4$ whatever its survival status and therefore does not widen the bounds.
In contrast, when the predicted survival probability is close to $0$ or $1$, the two candidate losses are close to $0$ and $1$, so their difference is close to $1$.
Thus, predictions farther from $1/2$ for individuals censored by $t$ contribute more to the width of the Brier score bounds.

We now consider bounds on the IBS over $[0,\tau]$.
Integrating the pointwise Brier score bounds over this period yields an outer set for the identified set of the IBS.
Since the inequalities in Proposition~\ref{prop:bs-bounds}(i) hold for every $t$ and every $Q\in\mathcal Q(P)$, integration gives
\begin{equation}
    \frac{1}{\tau}\int_0^\tau\underline{\operatorname{BS}}(t;P)\,dt
    \le
    \operatorname{IBS}(\tau;Q)
    \le
    \frac{1}{\tau}\int_0^\tau\overline{\operatorname{BS}}(t;P)\,dt.
    \label{eq:ibs-pointwise-bounds}
\end{equation}

Even when the pointwise bounds are sharp, the integrated upper bound need not be sharp.
At each time, the endpoints in~\eqref{eq:ibs-pointwise-bounds} choose for each censored individual the survival status that gives the smaller or the larger of the two candidate losses.
Under a single event time, the survival status $\mathbf 1(T>t)$ changes from one to zero at most once as $t$ increases and never returns to one.
Since $\widehat S(t\mid x)$ is nonincreasing in $t$, the loss of being event-free, $\{1-\widehat S(t\mid x)\}^2$, grows relative to the loss of having experienced the event, $\widehat S(t\mid x)^2$, as $t$ increases.
The ordering of the two losses therefore changes at most once, and so does each pointwise choice of survival status.
For the lower bound, the chosen status changes from one to zero, which a single event time produces or approaches.
For the upper bound, it increases from zero to one: the individual has experienced the event at earlier times and is event-free at later times.
No event time produces this sequence when both statuses are chosen on periods of positive length after censoring.
Therefore, the integrated lower bound is sharp, whereas the sharp upper bound must be obtained by choosing a single event time for each censored individual.

The observed data determine each individual's contribution to the IBS except for individuals censored before $\tau$.
For individuals with $Y\ge\tau$ or $\Delta=1$, this contribution is $L(\min(Y,\tau),X;\tau)/\tau$, with $L$ defined in~\eqref{eq:individual-loss}.
For individuals with $Y<\tau$ and $\Delta=0$, it can be $L(r,X;\tau)/\tau$ for any $r\in(Y,\tau]$, where $r=\min(T,\tau)$.
The sharp bounds therefore require the smallest and largest values of $L(r,X;\tau)$ over $r\in(Y,\tau]$ for each censored individual.

These smallest and largest values are determined by where the predicted survival probability crosses $1/2$.
At time $t$, the two candidate losses differ by $\{1-\widehat S(t\mid x)\}^2-\widehat S(t\mid x)^2=1-2\widehat S(t\mid x)$, so being event-free gives the smaller loss exactly when $\widehat S(t\mid x)>1/2$.
For $0\le y<\tau$ and covariate value $x$, the first time after $y$ at which the predicted survival probability falls to $1/2$ is
\[
    r^*(y,x)=\inf\bigl\{s\in[y,\tau]:\widehat S(s\mid x)\le1/2\bigr\},
\]
with $r^*(y,x)=\tau$ if this set is empty.
An event at $r^*(Y,X)$ gives the smaller loss at every time after censoring, so the loss of a censored individual is smallest at $r=r^*(Y,X)$.
The loss $L(r,X;\tau)$ increases in $r$ at the rate $1-2\widehat S(r\mid X)$, which is nondecreasing, so it is convex in $r$ and largest as $r\downarrow Y$ or at $r=\tau$.
Averaging these smallest and largest losses with the known contributions yields the following sharp bounds.

\begin{theoremEnd}{proposition}\label{prop:ibs-bounds}
    Define
    \begin{align*}
        \underline{\operatorname{IBS}}(\tau;P)
        =
        \frac{1}{\tau}\mathbb E_P\Bigl[
         & \mathbf 1(Y\ge\tau\text{ or }\Delta=1)\,L(\min(Y,\tau),X;\tau) \\
         & +\mathbf 1(Y<\tau,\Delta=0)\,L\bigl(r^*(Y,X),X;\tau\bigr)
            \Bigr],
    \end{align*}
    \begin{align*}
        \overline{\operatorname{IBS}}(\tau;P)
        =
        \frac{1}{\tau}\mathbb E_P\Bigl[
         & \mathbf 1(Y\ge\tau\text{ or }\Delta=1)\,L(\min(Y,\tau),X;\tau)          \\
         & +\mathbf 1(Y<\tau,\Delta=0)\max\bigl\{L(Y,X;\tau),L(\tau,X;\tau)\bigr\}
            \Bigr],
    \end{align*}
    and, for $0\le y<\tau$ and $x$,
    \[
        g(y,x)
        =
        \min\Bigl\{
        \int_y^\tau\bigl\{2\widehat S(t\mid x)-1\bigr\}_+\,dt,\
        \int_y^\tau\bigl\{1-2\widehat S(t\mid x)\bigr\}_+\,dt
        \Bigr\},
    \]
    where $a_+=\max\{a,0\}$.
    Then the following hold.
    \begin{enumerate}[label=(\roman*)]
        \item \emph{Outer set.}
              The interval $[\underline{\operatorname{IBS}}(\tau;P),\overline{\operatorname{IBS}}(\tau;P)]$ is an outer set for the identified set of $\operatorname{IBS}(\tau;Q)$; that is,
              \[
                  \underline{\operatorname{IBS}}(\tau;P)\le\operatorname{IBS}(\tau;Q)\le\overline{\operatorname{IBS}}(\tau;P)
                  \quad\text{for every }Q\in\mathcal Q(P).
              \]
        \item \emph{Sharpness.}
              $\underline{\operatorname{IBS}}(\tau;P)$ and $\overline{\operatorname{IBS}}(\tau;P)$ are the sharp lower and upper bounds on $\operatorname{IBS}(\tau;Q)$, but they need not be attained.
              That is, the closure of the identified set of $\operatorname{IBS}(\tau;Q)$ is $[\underline{\operatorname{IBS}}(\tau;P),\overline{\operatorname{IBS}}(\tau;P)]$.
        \item \emph{Width.}
              The width of the bounds is
              \begin{align*}
                  \overline{\operatorname{IBS}}(\tau;P)-\underline{\operatorname{IBS}}(\tau;P)
                  =
                  \frac{1}{\tau}\mathbb E_P\Bigl[
                   & \mathbf 1(Y<\tau,\Delta=0)                                                                           \\
                   & \times\Bigl\{\max\bigl\{L(Y,X;\tau),L(\tau,X;\tau)\bigr\}-L\bigl(r^*(Y,X),X;\tau\bigr)\Bigr\}\Bigr].
              \end{align*}
        \item \emph{Comparison with the integrated pointwise bounds.}
              \[
                  \frac{1}{\tau}\int_0^\tau\underline{\operatorname{BS}}(t;P)\,dt
                  =
                  \underline{\operatorname{IBS}}(\tau;P)
                  \le
                  \overline{\operatorname{IBS}}(\tau;P)
                  \le
                  \frac{1}{\tau}\int_0^\tau\overline{\operatorname{BS}}(t;P)\,dt,
              \]
              and the difference in the last inequality is
              \[
                  \frac{1}{\tau}\int_0^\tau\overline{\operatorname{BS}}(t;P)\,dt
                  -\overline{\operatorname{IBS}}(\tau;P)
                  =
                  \frac{1}{\tau}\mathbb E_P\bigl[\mathbf 1(Y<\tau,\Delta=0)\,g(Y,X)\bigr].
              \]
              In particular, the last inequality is strict if and only if $P\bigl(Y<\tau,\ \Delta=0,\ g(Y,X)>0\bigr)>0$.
    \end{enumerate}
\end{theoremEnd}
\begin{proofEnd}
    \emph{Preliminaries.}
    We first establish three properties of the loss $L$ for a censored individual.
    By~\eqref{eq:individual-loss}, for every $x$ and $0\le s\le r\le\tau$,
    \[
        L(r,x;\tau)-L(s,x;\tau)=\int_s^r\bigl\{1-2\widehat S(t\mid x)\bigr\}\,dt,
    \]
    since $\{1-\widehat S(t\mid x)\}^2-\widehat S(t\mid x)^2=1-2\widehat S(t\mid x)$.
    This identity uses only the definition of $L$.
    The integrand is bounded in absolute value by one, so $|L(r,x;\tau)-L(s,x;\tau)|\le|r-s|$.
    Since $\widehat S(t\mid x)$ is nonincreasing in $t$, the integrand is nondecreasing, so its integral, and therefore $r\mapsto L(r,x;\tau)$, is convex on $[0,\tau]$.

    Fix $0\le y<\tau$ and $x$, and write $r^*=r^*(y,x)$.
    By~\eqref{eq:individual-loss}, for $r\in[y,\tau]$,
    \begin{align*}
        L(r,x;\tau)
         & =\int_0^y\bigl\{1-\widehat S(t\mid x)\bigr\}^2\,dt+\int_y^r\bigl\{1-\widehat S(t\mid x)\bigr\}^2\,dt+\int_r^\tau\widehat S(t\mid x)^2\,dt \\
         & \ge\int_0^y\bigl\{1-\widehat S(t\mid x)\bigr\}^2\,dt+\int_y^\tau\min\bigl\{\widehat S(t\mid x)^2,\bigl\{1-\widehat S(t\mid x)\bigr\}^2\bigr\}\,dt.
    \end{align*}
    Since $\widehat S(t\mid x)$ is nonincreasing in $t$, $\widehat S(t\mid x)>1/2$ for $t\in[y,r^*)$ and $\widehat S(t\mid x)\le1/2$ for $t\in(r^*,\tau]$.
    Hence the minimum in the display equals $\{1-\widehat S(t\mid x)\}^2$ for $t\in[y,r^*)$ and $\widehat S(t\mid x)^2$ for $t\in(r^*,\tau]$.
    The single point $r^*$ does not affect the integrals, so taking $r=r^*$ attains the lower bound in the display:
    \begin{equation}
        \begin{aligned}
            L(r^*,x;\tau)
             & =\int_0^y\bigl\{1-\widehat S(t\mid x)\bigr\}^2\,dt+\int_y^\tau\min\bigl\{\widehat S(t\mid x)^2,\bigl\{1-\widehat S(t\mid x)\bigr\}^2\bigr\}\,dt \\
             & \le L(r,x;\tau),
            \qquad
            r\in[y,\tau].
        \end{aligned}
        \label{eq:proof-ibs-lower-loss}
    \end{equation}
    By convexity, $L(r,x;\tau)\le\max\{L(y,x;\tau),L(\tau,x;\tau)\}$ for every $r\in[y,\tau]$.

    These quantities are measurable functions of the observed data.
    By Tonelli's theorem and the joint measurability of $\widehat S$, the functions $L(Y,X;\tau)$, $L(\tau,X;\tau)$, and the integral expression in~\eqref{eq:proof-ibs-lower-loss} with $y=Y$ are measurable.
    For $r^*(Y,X)$, fix $c\le\tau$.
    By the definition of $r^*$, $r^*(Y,X)<c$ holds exactly when $\widehat S(s\mid X)\le1/2$ for some $s\in[Y,c)$.
    If such an $s$ exceeds $Y$, every rational $q\in(s,c)$ also satisfies $\widehat S(q\mid X)\le\widehat S(s\mid X)\le1/2$ because $\widehat S(t\mid X)$ is nonincreasing in $t$.
    Hence
    \[
        \{r^*(Y,X)<c\}
        =
        \bigl\{Y<c,\ \widehat S(Y\mid X)\le1/2\bigr\}
        \cup
        \bigcup_{q\in\mathbb Q}\bigl\{Y<q<c,\ \widehat S(q\mid X)\le1/2\bigr\}.
    \]
    Each event on the right-hand side is measurable by the joint measurability of $\widehat S$, so this countable union is measurable.
    For $c>\tau$, the event $\{r^*(Y,X)<c\}$ contains every point with $Y<\tau$, so $r^*(Y,X)$ is measurable on $Y<\tau$.
    Write $\underline\ell(O)$ and $\overline\ell(O)$ for $\tau^{-1}$ times the expressions inside the expectations defining $\underline{\operatorname{IBS}}(\tau;P)$ and $\overline{\operatorname{IBS}}(\tau;P)$, so that $\underline{\operatorname{IBS}}(\tau;P)=\mathbb E_P\,\underline\ell(O)$ and $\overline{\operatorname{IBS}}(\tau;P)=\mathbb E_P\,\overline\ell(O)$.

    \emph{Part (i).}
    Let $Q\in\mathcal Q(P)$.
    By compatibility, the observed data $O=(Y,\Delta,X)$ have distribution $P$ under $Q$, and $\operatorname{IBS}(\tau;Q)=\tau^{-1}\mathbb E_Q L(\min(T,\tau),X;\tau)$ by Section~\ref{subsec:framework}.
    If $\Delta=1$, then $T=Y$; if $\Delta=0$ and $Y\ge\tau$, then $T>Y\ge\tau$.
    In both cases, $\min(T,\tau)=\min(Y,\tau)$, so both $\underline\ell(O)$ and $\overline\ell(O)$ equal $\tau^{-1}L(\min(T,\tau),X;\tau)$.
    If $Y<\tau$ and $\Delta=0$, then $T>Y$, so $\min(T,\tau)\in(Y,\tau]$, and the preliminaries give $\underline\ell(O)\le\tau^{-1}L(\min(T,\tau),X;\tau)\le\overline\ell(O)$.
    Since $\underline\ell$ and $\overline\ell$ are functions of $O$ alone, their expectations under $Q$ equal those under $P$.
    Taking expectations gives $\underline{\operatorname{IBS}}(\tau;P)\le\operatorname{IBS}(\tau;Q)\le\overline{\operatorname{IBS}}(\tau;P)$.

    \emph{Part (ii).}
    Fix $\varepsilon>0$.
    On $Y<\tau$, define $T_\varepsilon(O)=\max\{r^*(Y,X),\min(Y+\varepsilon,\tau)\}$, a measurable function of $O$.
    Since $r^*(Y,X)\le\tau$ and $\min(Y+\varepsilon,\tau)>Y$, $T_\varepsilon(O)\in(Y,\tau]$.
    If $r^*(Y,X)\ge\min(Y+\varepsilon,\tau)$, then $T_\varepsilon(O)=r^*(Y,X)$; otherwise $r^*(Y,X)<T_\varepsilon(O)\le Y+\varepsilon\le r^*(Y,X)+\varepsilon$.
    Thus $|T_\varepsilon(O)-r^*(Y,X)|\le\varepsilon$, and the Lipschitz bound gives $L(T_\varepsilon(O),X;\tau)\le L(r^*(Y,X),X;\tau)+\varepsilon$.

    Given $O$, define $C=Y$ and
    \[
        T=
        \begin{cases}
            Y,                & \Delta=1,            \\
            Y+1,              & \Delta=0,\ Y\ge\tau, \\
            T_\varepsilon(O), & \Delta=0,\ Y<\tau,
        \end{cases}
    \]
    and let $Q_\varepsilon$ be the distribution of $(T,C,X)$ when $O$ follows $P$.
    In every case, $T\le C$ if and only if $\Delta=1$, and $\min(T,C)=Y$; hence $Q_\varepsilon\in\mathcal Q(P)$.
    Under $Q_\varepsilon$, $\min(T,\tau)=\min(Y,\tau)$ in the first two cases and $\min(T,\tau)=T_\varepsilon(O)$ in the third.
    Taking expectations under $P$ therefore gives $\operatorname{IBS}(\tau;Q_\varepsilon)\le\underline{\operatorname{IBS}}(\tau;P)+\varepsilon/\tau$.
    Together with part (i), the arbitrariness of $\varepsilon$ identifies $\underline{\operatorname{IBS}}(\tau;P)$ as the infimum of the identified set.

    For the upper bound, on $Y<\tau$ set $\overline T_\varepsilon(O)=\tau$ if $L(\tau,X;\tau)\ge L(Y,X;\tau)$ and $\overline T_\varepsilon(O)=\min(Y+\varepsilon,\tau)$ otherwise.
    This choice is measurable, lies in $(Y,\tau]$, and by the Lipschitz bound satisfies $L(\overline T_\varepsilon(O),X;\tau)\ge\max\{L(Y,X;\tau),L(\tau,X;\tau)\}-\varepsilon$.
    Replacing $T_\varepsilon(O)$ by $\overline T_\varepsilon(O)$ in the construction gives $\overline Q_\varepsilon\in\mathcal Q(P)$ with $\operatorname{IBS}(\tau;\overline Q_\varepsilon)\ge\overline{\operatorname{IBS}}(\tau;P)-\varepsilon/\tau$.
    Together with part (i), this identifies $\overline{\operatorname{IBS}}(\tau;P)$ as the supremum of the identified set.

    As in the proof of Proposition~\ref{prop:bs-bounds}(ii), any mixture of two distributions in $\mathcal Q(P)$ also induces $P$ and belongs to $\mathcal Q(P)$.
    Since the IBS is the expectation of a fixed individual loss, the IBS of the mixture is the same mixture of their IBS values.
    Thus every value between two attainable IBS values is also attainable, so the identified set is an interval.
    Its infimum and supremum established above therefore give its closure as $[\underline{\operatorname{IBS}}(\tau;P),\overline{\operatorname{IBS}}(\tau;P)]$.

    To establish the claim in part (ii) that the bounds need not be attained, we give two examples.
    The lower bound is unattained in the first example, and the upper bound is unattained in the second.
    Take $\tau=1$ and let $P(Y=0,\Delta=0,X=x_0)=1$ for a fixed covariate value $x_0$.
    Compatibility is then equivalent to $C=0$, $T>0$, and $X=x_0$ almost surely.
    Thus every compatible $Q$ has $m=\mathbb E_Q\min(T,1)\in(0,1]$: a strictly positive random variable cannot have expectation zero.
    Conversely, every $m\in(0,1]$ is realized by setting $C=0$ and $T=m$ deterministically.

    Both constant predictions below are nonincreasing in $t$.
    If $\widehat S\equiv0$, definition~\eqref{eq:individual-loss} gives $L(r,x_0;1)=r$ and $\operatorname{IBS}(1;Q)=m$.
    If $\widehat S\equiv1$, it gives $L(r,x_0;1)=1-r$ and $\operatorname{IBS}(1;Q)=1-m$.
    Hence the two identified sets are, respectively,
    \[
        \bigl\{\operatorname{IBS}(1;Q):Q\in\mathcal Q(P)\bigr\}
        =
        \begin{cases}
            (0,1], & \widehat S\equiv0, \\
            [0,1), & \widehat S\equiv1.
        \end{cases}
    \]
    The lower bound $0$ in the first example and the upper bound $1$ in the second are therefore unattained under every compatible $Q$.
    They are approached by $C=0$, $T=1/n$ as the positive integer $n$ tends to infinity.

    \emph{Part (iii).}
    The definitions of $\underline{\operatorname{IBS}}(\tau;P)$ and $\overline{\operatorname{IBS}}(\tau;P)$ differ only in the term for $Y<\tau$ and $\Delta=0$, and subtracting them gives the stated width.

    \emph{Part (iv).}
    By the definition in Proposition~\ref{prop:bs-bounds} and Tonelli's theorem, $\int_0^\tau\underline{\operatorname{BS}}(t;P)\,dt$ is the expectation under $P$ of
    \begin{align*}
        \int_0^\tau\Bigl[
         & \mathbf 1(Y>t)\bigl\{1-\widehat S(t\mid X)\bigr\}^2+\mathbf 1(Y\le t,\Delta=1)\,\widehat S(t\mid X)^2 \\
         & +\mathbf 1(Y\le t,\Delta=0)\min\bigl\{\widehat S(t\mid X)^2,\bigl\{1-\widehat S(t\mid X)\bigr\}^2\bigr\}
            \Bigr]dt.
    \end{align*}
    Tonelli's theorem applies because the integrand is nonnegative and measurable in $(t,O)$.
    If $Y\ge\tau$, only the first term remains on $[0,\tau)$, so this integral equals $L(\tau,X;\tau)$ by~\eqref{eq:individual-loss}.
    If $Y<\tau$ and $\Delta=1$, the first term remains on $[0,Y)$ and the second on $[Y,\tau]$, so the integral equals $L(Y,X;\tau)$.
    If $Y<\tau$ and $\Delta=0$, the first term remains on $[0,Y)$ and the minimum on $[Y,\tau]$, so the integral is the middle expression in~\eqref{eq:proof-ibs-lower-loss} with $y=Y$ and $x=X$, which equals $L(r^*(Y,X),X;\tau)$.
    These three cases agree with the expression defining $\underline{\operatorname{IBS}}(\tau;P)$, which proves the equality in part (iv).

    The same calculation for $\overline{\operatorname{BS}}$, with the minimum replaced by the maximum, changes only the contribution for $Y<\tau$ and $\Delta=0$.
    That contribution becomes the integral of $\{1-\widehat S(t\mid X)\}^2$ over $[0,Y)$ plus the integral of the larger of the two candidate losses over $[Y,\tau]$.
    The corresponding term of $\overline{\operatorname{IBS}}(\tau;P)$ is $\max\{L(Y,X;\tau),L(\tau,X;\tau)\}$.
    By~\eqref{eq:individual-loss}, $L(Y,X;\tau)$ and $L(\tau,X;\tau)$ share the same integral over $[0,Y)$ and integrate $\widehat S(t\mid X)^2$ and $\{1-\widehat S(t\mid X)\}^2$, respectively, over $[Y,\tau]$.
    Since the two candidate losses differ by $\{1-\widehat S(t\mid X)\}^2-\widehat S(t\mid X)^2=1-2\widehat S(t\mid X)$, the larger one exceeds $\widehat S(t\mid X)^2$ by $\{1-2\widehat S(t\mid X)\}_+$ and exceeds $\{1-\widehat S(t\mid X)\}^2$ by $\{2\widehat S(t\mid X)-1\}_+$.
    Integrating over $[Y,\tau]$, the integrated pointwise contribution therefore exceeds $L(Y,X;\tau)$ by $\int_Y^\tau\{1-2\widehat S(t\mid X)\}_+\,dt$ and exceeds $L(\tau,X;\tau)$ by $\int_Y^\tau\{2\widehat S(t\mid X)-1\}_+\,dt$.
    Since $c-\max\{u,v\}=\min\{c-u,c-v\}$ for real $c,u,v$, the difference between the two contributions is the smaller of these two integrals, which is $g(Y,X)$.

    Multiplying by $\mathbf 1(Y<\tau,\Delta=0)$, taking expectations under $P$, and dividing by $\tau$ give the stated difference.
    The middle inequality in part (iv) is $\underline{\operatorname{IBS}}(\tau;P)\le\overline{\operatorname{IBS}}(\tau;P)$, which follows from~\eqref{eq:proof-ibs-lower-loss} with $r=\tau$.
    Since $g\ge0$, the stated difference gives the last inequality.
    The difference is $\tau^{-1}$ times the expectation of the nonnegative random variable $\mathbf 1(Y<\tau,\Delta=0)\,g(Y,X)$, so it is positive if and only if $P(Y<\tau,\Delta=0,g(Y,X)>0)>0$.
\end{proofEnd}

By part (iii), only individuals censored before $\tau$ can widen the IBS bounds.
Each such individual's contribution to the width is $1/\tau$ times the difference between its largest loss, as $r\downarrow Y$ or at $r=\tau$, and its smallest loss, at $r=r^*(Y,X)$.
The lower bound thus corresponds to an event at the time the predicted survival probability falls to $1/2$, when the prediction is most consistent with the realized survival status at every time.
The upper bound corresponds to an event immediately after censoring or no event before $\tau$, whichever makes the prediction worse.
Individuals with observed events or censoring at or after $\tau$ contribute the same known loss to both bounds.
Thus the sharp bounds on the IBS place each censored individual's event time at one of these points.

\subsection{Comparison of two models}\label{subsec:two-models}

To compare two models $j$ and $k$, we take as the target quantity the difference in their IBSs,
\[
    \operatorname{IBS}_{jk}(\tau;Q)
    =
    \operatorname{IBS}_j(\tau;Q)-\operatorname{IBS}_k(\tau;Q).
\]
Since a smaller IBS indicates a smaller prediction error, $\operatorname{IBS}_{jk}(\tau;Q)<0$ means that model $j$ predicts better than model $k$ under $Q$.
Exchanging the two models reverses the sign, $\operatorname{IBS}_{kj}(\tau;Q)=-\operatorname{IBS}_{jk}(\tau;Q)$.
The identified set of this difference is defined as in Section~\ref{subsec:framework}, with $\theta(Q)=\operatorname{IBS}_{jk}(\tau;Q)$.
The ranking of the two models is therefore determined by where this identified set lies relative to zero.

Combining the bounds on each model's IBS in Proposition~\ref{prop:ibs-bounds} yields an outer set for the identified set of the difference, but this outer set is generally not sharp.
Writing $\underline{\operatorname{IBS}}_j(\tau;P)$ and $\overline{\operatorname{IBS}}_j(\tau;P)$ for the bounds on the IBS of model $j$, every $Q\in\mathcal Q(P)$ satisfies
\begin{equation}
    \underline{\operatorname{IBS}}_j(\tau;P)-\overline{\operatorname{IBS}}_k(\tau;P)
    \le
    \operatorname{IBS}_{jk}(\tau;Q)
    \le
    \overline{\operatorname{IBS}}_j(\tau;P)-\underline{\operatorname{IBS}}_k(\tau;P).
    \label{eq:ibs-diff-separate-bounds}
\end{equation}
The lower end of this interval, however, minimizes the IBS of model $j$ and maximizes that of model $k$ with separate event times, whereas each individual has a single event time shared by both models.
In the extreme case in which the two models output the same predictions, $\operatorname{IBS}_{jk}(\tau;Q)=0$ for every $Q$, so the identified set is $\{0\}$.
The interval in~\eqref{eq:ibs-diff-separate-bounds} nevertheless has width $2\{\overline{\operatorname{IBS}}_j(\tau;P)-\underline{\operatorname{IBS}}_j(\tau;P)\}$, which is positive whenever the IBS of model $j$ is not point identified.
Hence, the sharp bounds must be obtained by minimizing and maximizing the difference in losses itself under a single event time common to both models.

To do so, we assign one value of $r$ to each censored individual and use it for both models.
Let
\[
    D_{jk}(r,x)=L_j(r,x;\tau)-L_k(r,x;\tau),
    \qquad
    0\le r\le\tau,
\]
denote the difference in the losses of the two models for an individual whose event time truncated at $\tau$ is $r$ and whose covariates are $x$.
By the representation in Section~\ref{subsec:framework}, $\operatorname{IBS}_{jk}(\tau;Q)=\tau^{-1}\mathbb E_Q D_{jk}(\min(T,\tau),X)$.
As in Section~\ref{subsec:single-model}, the data determine $\min(T,\tau)=\min(Y,\tau)$ unless the individual is censored before $\tau$, in which case $\min(T,\tau)$ can take any value in $(Y,\tau]$.
The sharp bounds therefore require the smallest and largest values of $D_{jk}(r,X)$ over $r\in(Y,\tau]$ for each censored individual.

The smallest and largest values depend on where the two predicted survival curves cross.
As in the proof of Proposition~\ref{prop:ibs-bounds}, the definition of $L_j$ in~\eqref{eq:individual-loss} directly gives the following identity:
\[
    L_j(r,x;\tau)-L_j(y,x;\tau)=\int_y^r\bigl\{1-2\widehat S_j(t\mid x)\bigr\}\,dt,
    \qquad
    0\le y\le r\le\tau.
\]
Taking the difference between the two models gives
\begin{equation}
    D_{jk}(r,x)
    =
    D_{jk}(y,x)
    +2\int_y^r\bigl\{\widehat S_k(t\mid x)-\widehat S_j(t\mid x)\bigr\}\,dt,
    \qquad
    0\le y\le r\le\tau.
    \label{eq:loss-diff-increment}
\end{equation}
Hence $D_{jk}(r,X)$ decreases in $r$ where model $j$ predicts a higher survival probability than model $k$, and increases where it predicts a lower one.
For example, if $\widehat S_j(t\mid X)\ge\widehat S_k(t\mid X)$ for all $t\in[Y,\tau]$, the infimum over $r\in(Y,\tau]$ is attained at $r=\tau$, and the supremum is approached as $r\downarrow Y$.
In general, the infimum and supremum for a censored individual are attained or approached at $r\downarrow Y$, at $r=\tau$, or at a point where $\widehat S_j(t\mid X)-\widehat S_k(t\mid X)$ changes sign.
Unlike the loss of a single model, $D_{jk}(r,X)$ need not be convex in $r$, so these extremes do not reduce to a closed form.
We therefore define the individual contributions to the lower and upper bounds by
\begin{align*}
    \underline d_{jk}(O)
     & =
    \frac{1}{\tau}\Bigl\{
    \mathbf 1(Y\ge\tau\text{ or }\Delta=1)\,D_{jk}(\min(Y,\tau),X)
    +\mathbf 1(Y<\tau,\Delta=0)\inf_{r\in(Y,\tau]}D_{jk}(r,X)
    \Bigr\}, \\
    \overline d_{jk}(O)
     & =
    \frac{1}{\tau}\Bigl\{
    \mathbf 1(Y\ge\tau\text{ or }\Delta=1)\,D_{jk}(\min(Y,\tau),X)
    +\mathbf 1(Y<\tau,\Delta=0)\sup_{r\in(Y,\tau]}D_{jk}(r,X)
    \Bigr\},
\end{align*}
and set $\underline{\operatorname{IBS}}_{jk}(\tau;P)=\mathbb E_P\,\underline d_{jk}(O)$ and $\overline{\operatorname{IBS}}_{jk}(\tau;P)=\mathbb E_P\,\overline d_{jk}(O)$.
Since $D_{kj}=-D_{jk}$, exchanging the two models gives $\overline{\operatorname{IBS}}_{kj}(\tau;P)=-\underline{\operatorname{IBS}}_{jk}(\tau;P)$.

\begin{theoremEnd}{proposition}\label{prop:pair-bounds}
    For models $j$ and $k$, the following hold.
    \begin{enumerate}[label=(\roman*)]
        \item \emph{Outer set.}
              The interval $[\underline{\operatorname{IBS}}_{jk}(\tau;P),\overline{\operatorname{IBS}}_{jk}(\tau;P)]$ is an outer set for the identified set of $\operatorname{IBS}_{jk}(\tau;Q)$; that is,
              \[
                  \underline{\operatorname{IBS}}_{jk}(\tau;P)\le\operatorname{IBS}_{jk}(\tau;Q)\le\overline{\operatorname{IBS}}_{jk}(\tau;P)
                  \quad\text{for every }Q\in\mathcal Q(P).
              \]
        \item \emph{Sharpness.}
              The lower bound $\underline{\operatorname{IBS}}_{jk}(\tau;P)$ and the upper bound $\overline{\operatorname{IBS}}_{jk}(\tau;P)$ on $\operatorname{IBS}_{jk}(\tau;Q)$ are sharp, but they need not be attained.
              That is, the closure of the identified set of $\operatorname{IBS}_{jk}(\tau;Q)$ is
              \[
                  \bigl[\underline{\operatorname{IBS}}_{jk}(\tau;P),\overline{\operatorname{IBS}}_{jk}(\tau;P)\bigr].
              \]
        \item \emph{Width.}
              The width of the bounds is
              \[
                  \overline{\operatorname{IBS}}_{jk}(\tau;P)-\underline{\operatorname{IBS}}_{jk}(\tau;P)
                  =
                  \frac{1}{\tau}\mathbb E_P\Bigl[\mathbf 1(Y<\tau,\Delta=0)\Bigl\{\sup_{r\in(Y,\tau]}D_{jk}(r,X)-\inf_{r\in(Y,\tau]}D_{jk}(r,X)\Bigr\}\Bigr].
              \]
        \item \emph{Comparison with the combined bounds.}
              \[
                  \underline{\operatorname{IBS}}_j(\tau;P)-\overline{\operatorname{IBS}}_k(\tau;P)
                  \le
                  \underline{\operatorname{IBS}}_{jk}(\tau;P)
                  \le
                  \overline{\operatorname{IBS}}_{jk}(\tau;P)
                  \le
                  \overline{\operatorname{IBS}}_j(\tau;P)-\underline{\operatorname{IBS}}_k(\tau;P).
              \]
    \end{enumerate}
\end{theoremEnd}
\begin{proofEnd}
    \emph{Preliminaries.}
    The function $x\mapsto D_{jk}(r,x)$ is measurable because each model's prediction is measurable.
    Also, $|D_{jk}(r,x)|\le\tau$ because $0\le L_j(r,x;\tau),L_k(r,x;\tau)\le\tau$, so the expectations below are finite.
    By the preliminaries in the proof of Proposition~\ref{prop:ibs-bounds}, each model's loss is Lipschitz in $r$ with constant one, so $D_{jk}(r,x)$ is continuous in $r$.
    Unlike $L_j$, however, $D_{jk}$ need not be convex in $r$ even though both predictions are nonincreasing.
    We therefore approximate its infimum and supremum through a countable set of candidates.

    \emph{Part (i).}
    Let $Q\in\mathcal Q(P)$.
    Then $O$ follows $P$ under $Q$, and by the representation in Section~\ref{subsec:framework}, $\operatorname{IBS}_{jk}(\tau;Q)=\tau^{-1}\mathbb E_Q D_{jk}(\min(T,\tau),X)$.
    If $\Delta=1$, or if $\Delta=0$ and $Y\ge\tau$, then $\min(T,\tau)=\min(Y,\tau)$, so both $\underline d_{jk}(O)$ and $\overline d_{jk}(O)$ equal $\tau^{-1}D_{jk}(\min(T,\tau),X)$.
    If $Y<\tau$ and $\Delta=0$, then $\min(T,\tau)\in(Y,\tau]$, so $D_{jk}(\min(T,\tau),X)$ lies between the infimum and the supremum of $D_{jk}(r,X)$ over this interval.
    Hence, $Q$-almost surely,
    \[
        \underline d_{jk}(O)\le\frac{1}{\tau}D_{jk}(\min(T,\tau),X)\le\overline d_{jk}(O).
    \]
    Since $\underline d_{jk}$ and $\overline d_{jk}$ are functions of $O$ alone, taking expectations under $Q$ and replacing them with expectations under $P$ give part (i).

    \emph{Part (ii).}
    Fix $\varepsilon>0$.
    By continuity of $D_{jk}$ in $r$, the infimum over $(Y,\tau]$ can be computed using a countable set of candidates.
    Enumerate $(\mathbb Q\cap[0,\tau])\cup\{\tau\}$ as $q_1,q_2,\ldots$, and call a candidate $q_h$ admissible if $q_h>Y$.
    On $Y<\tau$, the admissible candidates are dense in $(Y,\tau]$ and include $\tau$, so every $r\in(Y,\tau]$ is approached by admissible candidates whose values of $D_{jk}$ converge to $D_{jk}(r,X)$.
    For each $h$, let $V_h=D_{jk}(q_h,X)$ if $q_h>Y$ and $V_h=+\infty$ otherwise.
    Each $V_h$ is measurable, and on $Y<\tau$,
    \[
        \inf_{r\in(Y,\tau]}D_{jk}(r,X)=\inf_{h\ge1}V_h,
    \]
    which is measurable as the infimum of countably many measurable functions and finite because $|D_{jk}|\le\tau$.
    By this equality, some $V_h$ is smaller than $\inf_{r\in(Y,\tau]}D_{jk}(r,X)+\varepsilon$; such a $V_h$ is finite, so its candidate $q_h$ is admissible.

    On $Y<\tau$, let $r_\varepsilon(O)$ be the first admissible $q_h$ such that $D_{jk}(q_h,X)<\inf_{r\in(Y,\tau]}D_{jk}(r,X)+\varepsilon$, and set $r_\varepsilon(O)=\tau$ on $Y\ge\tau$.
    The event that index $h$ is selected is the measurable event that $q_h$ satisfies this inequality and every earlier candidate fails it or is inadmissible, so $r_\varepsilon$ is measurable.
    Given $O$, define $C=Y$ and
    \[
        T=
        \begin{cases}
            Y,                & \Delta=1,            \\
            Y+1,              & \Delta=0,\ Y\ge\tau, \\
            r_\varepsilon(O), & \Delta=0,\ Y<\tau,
        \end{cases}
    \]
    and let $Q_\varepsilon$ be the distribution of $(T,C,X)$ when $O$ follows $P$.
    In every case, $T\le C$ if and only if $\Delta=1$, and $\min(T,C)=Y$; hence $Q_\varepsilon\in\mathcal Q(P)$.
    This single value of $T$ is shared by both models, and taking expectations under $P$ gives $\operatorname{IBS}_{jk}(\tau;Q_\varepsilon)\le\underline{\operatorname{IBS}}_{jk}(\tau;P)+\varepsilon/\tau$.
    Together with part (i), the arbitrariness of $\varepsilon$ identifies $\underline{\operatorname{IBS}}_{jk}(\tau;P)$ as the infimum of the identified set.

    For the upper bound, use the same candidates, assign $-\infty$ to inadmissible candidates, and choose the first admissible $q_h$ with $D_{jk}(q_h,X)>\sup_{r\in(Y,\tau]}D_{jk}(r,X)-\varepsilon$ on $Y<\tau$.
    The same measurability and compatibility arguments give $\overline Q_\varepsilon\in\mathcal Q(P)$ with $\operatorname{IBS}_{jk}(\tau;\overline Q_\varepsilon)\ge\overline{\operatorname{IBS}}_{jk}(\tau;P)-\varepsilon/\tau$, which identifies $\overline{\operatorname{IBS}}_{jk}(\tau;P)$ as the supremum.
    Mixtures of distributions compatible with $P$ still induce $P$, and the IBS difference of a mixture is the same mixture of their IBS differences.
    Hence the identified set is an interval, and its closure is the stated closed interval.

    \emph{Non-attainment examples.}
    To establish the claim in part (ii) that the bounds need not be attained, we give two examples.
    The lower bound is unattained in the first example, and the upper bound is unattained in the second.
    Take $\tau=1$, $P(Y=0,\Delta=0,X=x_0)=1$, and $\widehat S_k\equiv1/2$.
    Every compatible $Q$ has $C=0$, $T>0$, and $m=\mathbb E_Q\min(T,1)\in(0,1]$, and every such $m$ is realized by $T=m$, $C=0$.
    Definition~\eqref{eq:individual-loss} gives $L_k(r,x_0;1)=1/4$.
    Taking $\widehat S_j\equiv0$ gives $D_{jk}(r,x_0)=r-1/4$ and $\operatorname{IBS}_{jk}(1;Q)=m-1/4$.
    Taking $\widehat S_j\equiv1$ instead gives $D_{jk}(r,x_0)=3/4-r$ and $\operatorname{IBS}_{jk}(1;Q)=3/4-m$.
    Thus
    \[
        \bigl\{\operatorname{IBS}_{jk}(1;Q):Q\in\mathcal Q(P)\bigr\}
        =
        \begin{cases}
            (-1/4,3/4], & \widehat S_j\equiv0,\\
            [-1/4,3/4), & \widehat S_j\equiv1.
        \end{cases}
    \]
    The lower bound $-1/4$ in the first example and the upper bound $3/4$ in the second are not attained by any compatible $Q$, but are approached by $C=0$, $T=1/n$ as $n\to\infty$.

    \emph{Part (iii).}
    The definitions of $\underline d_{jk}(O)$ and $\overline d_{jk}(O)$ differ only in the term for $Y<\tau$ and $\Delta=0$, and subtracting their expectations gives the stated width.

    \emph{Part (iv).}
    By part (ii) and Proposition~\ref{prop:ibs-bounds}(ii) applied to models $j$ and $k$, and since $\inf(f-g)\ge\inf f-\sup g$ on a common domain,
    \begin{align*}
        \underline{\operatorname{IBS}}_{jk}(\tau;P)
         & =
        \inf_{Q\in\mathcal Q(P)}\bigl\{\operatorname{IBS}_j(\tau;Q)-\operatorname{IBS}_k(\tau;Q)\bigr\}                       \\
         & \ge
        \inf_{Q\in\mathcal Q(P)}\operatorname{IBS}_j(\tau;Q)-\sup_{Q\in\mathcal Q(P)}\operatorname{IBS}_k(\tau;Q)
        =
        \underline{\operatorname{IBS}}_j(\tau;P)-\overline{\operatorname{IBS}}_k(\tau;P).
    \end{align*}
    The inequality for the upper bound follows in the same way from $\sup(f-g)\le\sup f-\inf g$.
    The middle inequality follows directly from $\inf_{r\in(Y,\tau]}D_{jk}(r,X)\le\sup_{r\in(Y,\tau]}D_{jk}(r,X)$ in the individual contributions.
\end{proofEnd}

If the upper bound on the IBS difference is negative, model $j$ has a smaller IBS than model $k$ under every full-data distribution compatible with the observed data.
We give this situation a name.

\begin{definition}[Robust strict dominance]\label{def:robust-strict-dominance}
    We say that model $j$ robustly strictly dominates model $k$ if
    \[
        \overline{\operatorname{IBS}}_{jk}(\tau;P)<0.
    \]
\end{definition}

\noindent By Proposition~\ref{prop:pair-bounds}(i), robust strict dominance implies $\operatorname{IBS}_{jk}(\tau;Q)\le\overline{\operatorname{IBS}}_{jk}(\tau;P)<0$ for every $Q\in\mathcal Q(P)$, so model $j$ is better than model $k$ by at least $-\overline{\operatorname{IBS}}_{jk}(\tau;P)$ regardless of $Q$.
It is sufficient but not necessary for model $j$ to be strictly better under every $Q$, because the upper bound may equal zero without being attained.
If instead $\underline{\operatorname{IBS}}_{jk}(\tau;P)<0<\overline{\operatorname{IBS}}_{jk}(\tau;P)$, Proposition~\ref{prop:pair-bounds}(ii) implies that some compatible full-data distributions favor model $j$ and others favor model $k$.
In that case, the observed data alone cannot determine which model predicts better.

\subsection{Comparison of multiple models}\label{subsec:multiple-models}

When $K\ge3$ models are compared, whether model $j$ is the best is determined by the difference between its IBS and that of its best competitor,
\[
    M_j(\tau;Q)
    =
    \max_{k\ne j}\operatorname{IBS}_{jk}(\tau;Q)
    =
    \operatorname{IBS}_j(\tau;Q)-\min_{k\ne j}\operatorname{IBS}_k(\tau;Q).
\]
Since a smaller IBS is better, $M_j(\tau;Q)<0$ means that model $j$ predicts strictly better than every other model under $Q$.
The identified set of $M_j(\tau;Q)$ is defined as in Section~\ref{subsec:framework}, with $\theta(Q)=M_j(\tau;Q)$.
If this identified set lies below zero, model $j$ is the best under every full-data distribution compatible with the observed data.

The sharp upper bound on $M_j(\tau;Q)$ equals the largest of the upper bounds on the pairwise differences.
To make the maximum of these differences large, it suffices to make one of them large.
The supremum over compatible full-data distributions can therefore be interchanged with the maximum over the finitely many competitors.
The following proposition expresses the resulting upper bound using the pairwise bounds from Section~\ref{subsec:two-models}.

\begin{theoremEnd}{proposition}\label{prop:best-bounds}
    For a model $j$, define
    \[
        \overline M_j(\tau;P)
        =
        \max_{k\ne j}\overline{\operatorname{IBS}}_{jk}(\tau;P).
    \]
    Then $M_j(\tau;Q)\le\overline M_j(\tau;P)$ for every $Q\in\mathcal Q(P)$, and this upper bound is sharp:
    \[
        \sup_{Q\in\mathcal Q(P)}M_j(\tau;Q)=\overline M_j(\tau;P).
    \]
\end{theoremEnd}
\begin{proofEnd}
    By Proposition~\ref{prop:pair-bounds}(i), every $Q\in\mathcal Q(P)$ satisfies
    \[
        M_j(\tau;Q)
        =
        \max_{k\ne j}\operatorname{IBS}_{jk}(\tau;Q)
        \le
        \max_{k\ne j}\overline{\operatorname{IBS}}_{jk}(\tau;P)
        =
        \overline M_j(\tau;P).
    \]
    For each $k\ne j$, the inequality $M_j(\tau;Q)\ge\operatorname{IBS}_{jk}(\tau;Q)$ and Proposition~\ref{prop:pair-bounds}(ii) give
    \[
        \sup_{Q\in\mathcal Q(P)}M_j(\tau;Q)
        \ge
        \sup_{Q\in\mathcal Q(P)}\operatorname{IBS}_{jk}(\tau;Q)
        =
        \overline{\operatorname{IBS}}_{jk}(\tau;P).
    \]
    Maximizing the right-hand side over $k\ne j$ gives the reverse inequality and proves sharpness.
\end{proofEnd}

If the upper bound on $M_j(\tau;Q)$ is negative, model $j$ is the best under every full-data distribution compatible with the observed data.
We give this situation a name.

\begin{definition}[Robustly strictly best model]\label{def:robustly-strictly-best}
    We say that model $j$ is robustly strictly best among the $K$ models if
    \[
        \overline M_j(\tau;P)<0.
    \]
\end{definition}

\noindent By Proposition~\ref{prop:best-bounds}, model $j$ is robustly strictly best among the $K$ models if and only if it robustly strictly dominates every other model in the sense of Definition~\ref{def:robust-strict-dominance}.
As in the two-model case, this condition is sufficient but not necessary for model $j$ to be strictly best under every $Q$, because the upper bound may equal zero without being attained.
If $\overline M_j(\tau;P)>0$, sharpness implies that some compatible full-data distribution gives model $j$ a larger IBS than at least one competitor.
Thus certification of a robustly strictly best model requires only the upper bounds on the pairwise differences.

\section{Estimation and inference}\label{sec:estimation-inference}
\pratendSetLocal{category=inference}

This section develops estimation and inference for the sharp bounds derived in Section~\ref{sec:partial-identification}: both bounds for a single model and for a pairwise difference, and the upper bound for the difference from the best competitor.
Section~\ref{subsec:estimators} defines estimators of the bounds as sample averages over the evaluation sample, and Section~\ref{subsec:asymptotics} states the assumptions and establishes their consistency and asymptotic distributions for a fixed observed-data distribution.
Section~\ref{subsec:confidence-tests} then constructs confidence regions for a single model and a pairwise difference, together with tests of robust strict dominance and of whether a model is robustly strictly best among the $K$ models.

\subsection{Estimators}\label{subsec:estimators}

For fitted prediction models, we estimate the sharp bounds in Section~\ref{sec:partial-identification} using the evaluation sample $O_1,\ldots,O_n$.
For a single model and for a pairwise difference, each bound is the expectation of an individual contribution, so we replace that expectation with the corresponding sample average.
For the difference from the best competitor, we estimate the upper bound by the largest estimated pairwise upper bound.

For a single model, we estimate the bounds on the IBS by the sample averages of the individual contributions in Proposition~\ref{prop:ibs-bounds}:
\begin{align*}
    \widehat{\underline{\operatorname{IBS}}}(\tau)
     & =
    \frac{1}{n\tau}\sum_{i=1}^n\Bigl\{
    \mathbf 1(Y_i\ge\tau\text{ or }\Delta_i=1)\,L(\min(Y_i,\tau),X_i;\tau) \\
     & \qquad\qquad\qquad
    +\mathbf 1(Y_i<\tau,\Delta_i=0)\,L\bigl(r^*(Y_i,X_i),X_i;\tau\bigr)
    \Bigr\},                                                               \\
    \widehat{\overline{\operatorname{IBS}}}(\tau)
     & =
    \frac{1}{n\tau}\sum_{i=1}^n\Bigl\{
    \mathbf 1(Y_i\ge\tau\text{ or }\Delta_i=1)\,L(\min(Y_i,\tau),X_i;\tau) \\
     & \qquad\qquad\qquad
    +\mathbf 1(Y_i<\tau,\Delta_i=0)\max\bigl\{L(Y_i,X_i;\tau),L(\tau,X_i;\tau)\bigr\}
    \Bigr\}.
\end{align*}
Here $r^*(Y_i,X_i)$ is the infimum of the times $s\in[Y_i,\tau]$ such that $\widehat S(s\mid X_i)\le1/2$, with value $\tau$ if no such time exists, as defined in Section~\ref{subsec:single-model}.
By the nonincreasing property in Assumption~\ref{assump:survival}, the lower contribution of a censored individual is expressed using $r^*(Y_i,X_i)$, whereas the upper contribution is the larger of $L(Y_i,X_i;\tau)$ and $L(\tau,X_i;\tau)$.
When several models are compared, we write $\widehat{\underline{\operatorname{IBS}}}_j(\tau)$ and $\widehat{\overline{\operatorname{IBS}}}_j(\tau)$ for model $j$.
The estimated interval $[\widehat{\underline{\operatorname{IBS}}}(\tau),\widehat{\overline{\operatorname{IBS}}}(\tau)]$ is an estimate of the closure of the identified set of the IBS.

For two models $j$ and $k$, we estimate the bounds on the IBS difference by the sample averages of the individual contributions $\underline d_{jk}(O_i)$ and $\overline d_{jk}(O_i)$ in Section~\ref{subsec:two-models}, rather than by differences between the single-model estimators:
\[
    \widehat{\underline{\operatorname{IBS}}}_{jk}(\tau)
    =
    \frac{1}{n}\sum_{i=1}^n\underline d_{jk}(O_i),
    \qquad
    \widehat{\overline{\operatorname{IBS}}}_{jk}(\tau)
    =
    \frac{1}{n}\sum_{i=1}^n\overline d_{jk}(O_i).
\]
Since $\overline d_{kj}=-\underline d_{jk}$, exchanging the two models gives $\widehat{\overline{\operatorname{IBS}}}_{kj}(\tau)=-\widehat{\underline{\operatorname{IBS}}}_{jk}(\tau)$.
The differences $\widehat{\underline{\operatorname{IBS}}}_j(\tau)-\widehat{\overline{\operatorname{IBS}}}_k(\tau)$ and $\widehat{\overline{\operatorname{IBS}}}_j(\tau)-\widehat{\underline{\operatorname{IBS}}}_k(\tau)$ would instead estimate the wider interval in~\eqref{eq:ibs-diff-separate-bounds}.
In $\underline d_{jk}(O_i)$ and $\overline d_{jk}(O_i)$, the same value of $r$ enters the losses of both models.
Thus the estimators share the event time of each censored individual between the two models, as the sharp bounds do.

For the difference between model $j$ and its best competitor, we estimate the upper bound by the largest estimated pairwise upper bound, following Proposition~\ref{prop:best-bounds}:
\[
    \widehat{\overline M}_j(\tau)=\max_{k\ne j}\widehat{\overline{\operatorname{IBS}}}_{jk}(\tau).
\]
The individual contributions for each pair have already been computed, so this estimator requires only taking their largest sample average.
No additional optimization over event times is needed.
Since this estimator is a maximum of sample averages rather than a sample average itself, Section~\ref{subsec:asymptotics} treats its asymptotic distribution separately.

\subsection{Asymptotic properties}\label{subsec:asymptotics}

We establish the asymptotic properties of the estimators under the following assumptions.

\begin{assumption}[Independent evaluation sample]\label{assump:iid}
    The evaluation sample $O_1,\ldots,O_n$ consists of independent and identically distributed copies of $O=(Y,\Delta,X)$ with a fixed observed-data distribution $P$ that does not depend on $n$.
\end{assumption}

\begin{assumption}[Predicted survival functions]\label{assump:survival}
    For each model $j$, the predicted survival function $\widehat S_j(t\mid x)$ is measurable in $(t,x)$ and takes values in $[0,1]$.
    For every $x\in\mathcal X$, it is nonincreasing in time: $\widehat S_j(t_1\mid x)\ge\widehat S_j(t_2\mid x)$ whenever $0\le t_1\le t_2\le\tau$.
\end{assumption}

\begin{assumption}[Independent model construction]\label{assump:independent-training}
    All data used to construct the prediction models, including preprocessing, training, and hyperparameter tuning, are independent of the evaluation sample.
    The fitted prediction models are held fixed during evaluation.
\end{assumption}

\begin{assumption}[Prespecified comparison]\label{assump:prespecified}
    The candidate model $j$, the set of $K$ compared models, and the evaluation horizon $\tau$ are fixed before the evaluation sample is examined.
\end{assumption}

\noindent Assumption~\ref{assump:survival} formalizes the properties of the predicted survival functions introduced in Section~\ref{subsec:framework}.
Under Assumption~\ref{assump:independent-training}, inference is conditional on the data used to construct the models, so the fitted prediction functions are treated as fixed.
A validation sample may be used for tuning if it is independent of the evaluation sample.
The results therefore describe the sampling variability of the evaluation sample, conditional on the fitted models.

For model $j$, let $\underline\ell_j(O)$ and $\overline\ell_j(O)$ denote the summands of $\widehat{\underline{\operatorname{IBS}}}_j(\tau)$ and $\widehat{\overline{\operatorname{IBS}}}_j(\tau)$ in Section~\ref{subsec:estimators}, so that, for example, $\widehat{\underline{\operatorname{IBS}}}_j(\tau)=n^{-1}\sum_{i=1}^n\underline\ell_j(O_i)$ and $\underline{\operatorname{IBS}}_j(\tau;P)=\mathbb E_P\,\underline\ell_j(O)$.
Since $0\le L_j\le\tau$, the contributions $\underline\ell_j$ and $\overline\ell_j$ take values in $[0,1]$, and $\underline d_{jk}$ and $\overline d_{jk}$ take values in $[-1,1]$.
The law of large numbers and the central limit theorem therefore apply without further moment conditions.

The estimators of the bounds all converge almost surely to the corresponding bounds.

\begin{theoremEnd}{proposition}[Strong consistency]\label{prop:consistency}
    Under Assumptions~\ref{assump:iid}, \ref{assump:survival}, and~\ref{assump:independent-training}, the following hold as $n\to\infty$.
    \begin{enumerate}[label=(\roman*)]
        \item For each model $j$, $\widehat{\underline{\operatorname{IBS}}}_j(\tau)\xrightarrow{\mathrm{a.s.}}\underline{\operatorname{IBS}}_j(\tau;P)$ and $\widehat{\overline{\operatorname{IBS}}}_j(\tau)\xrightarrow{\mathrm{a.s.}}\overline{\operatorname{IBS}}_j(\tau;P)$.
        \item For each pair of models $j$ and $k$, $\widehat{\underline{\operatorname{IBS}}}_{jk}(\tau)\xrightarrow{\mathrm{a.s.}}\underline{\operatorname{IBS}}_{jk}(\tau;P)$ and $\widehat{\overline{\operatorname{IBS}}}_{jk}(\tau)\xrightarrow{\mathrm{a.s.}}\overline{\operatorname{IBS}}_{jk}(\tau;P)$.
        \item For each model $j$, $\widehat{\overline M}_j(\tau)\xrightarrow{\mathrm{a.s.}}\overline M_j(\tau;P)$.
    \end{enumerate}
\end{theoremEnd}
\begin{proofEnd}
    \emph{Parts (i) and (ii).}
    The individual contributions are measurable and bounded, as established in Propositions~\ref{prop:ibs-bounds} and~\ref{prop:pair-bounds}.
    Each estimator is the sample average of independent and identically distributed copies of the corresponding contribution.
    Its expectation is the bound being estimated, so the strong law of large numbers proves parts (i) and (ii).

    \emph{Part (iii).}
    By the definitions of the upper bound and its estimator,
    \[
        \bigl|\widehat{\overline M}_j(\tau)-\overline M_j(\tau;P)\bigr|
        \le
        \max_{k\ne j}
        \bigl|\widehat{\overline{\operatorname{IBS}}}_{jk}(\tau)
        -\overline{\operatorname{IBS}}_{jk}(\tau;P)\bigr|.
    \]
    Because the number of competitors is finite, part (ii) makes the right-hand side converge to zero almost surely.
\end{proofEnd}
Part (iii) does not require that the competitor attaining the upper bound be unique.
The proposition concerns the bounds, not the IBS itself, because the IBS is only known to lie in its identified set.
Increasing the sample size removes the estimation error, but the width of the identified set remains.

The estimators of the bounds for a single model and for two models are sample averages, so the lower and upper estimators are jointly asymptotically normal.

\begin{theoremEnd}{theorem}[Joint asymptotic normality]\label{thm:joint-normality}
    Under Assumptions~\ref{assump:iid}, \ref{assump:survival}, and~\ref{assump:independent-training}, the following hold as $n\to\infty$.
    \begin{enumerate}[label=(\roman*)]
        \item For each model $j$,
              \[
                  \sqrt n
                  \begin{pmatrix}
                      \widehat{\underline{\operatorname{IBS}}}_j(\tau)-\underline{\operatorname{IBS}}_j(\tau;P) \\
                      \widehat{\overline{\operatorname{IBS}}}_j(\tau)-\overline{\operatorname{IBS}}_j(\tau;P)
                  \end{pmatrix}
                  \xrightarrow{d}
                  N_2(0,\Sigma_j),
                  \qquad
                  \Sigma_j=\operatorname{Var}_P
                  \begin{pmatrix}
                      \underline\ell_j(O) \\
                      \overline\ell_j(O)
                  \end{pmatrix}.
              \]
        \item For each pair of models $j$ and $k$,
              \[
                  \sqrt n
                  \begin{pmatrix}
                      \widehat{\underline{\operatorname{IBS}}}_{jk}(\tau)-\underline{\operatorname{IBS}}_{jk}(\tau;P) \\
                      \widehat{\overline{\operatorname{IBS}}}_{jk}(\tau)-\overline{\operatorname{IBS}}_{jk}(\tau;P)
                  \end{pmatrix}
                  \xrightarrow{d}
                  N_2(0,\Sigma_{jk}),
                  \qquad
                  \Sigma_{jk}=\operatorname{Var}_P
                  \begin{pmatrix}
                      \underline d_{jk}(O) \\
                      \overline d_{jk}(O)
                  \end{pmatrix}.
              \]
        \item The sample covariance matrices of the contributions over $O_1,\ldots,O_n$ converge almost surely to $\Sigma_j$ and $\Sigma_{jk}$.
    \end{enumerate}
    The covariance matrices $\Sigma_j$ and $\Sigma_{jk}$ may be singular, in which case the limiting normal distributions are degenerate.
\end{theoremEnd}
\begin{proofEnd}
    The contributions are bounded, as noted at the beginning of Section~\ref{subsec:asymptotics}, so they have finite second moments.
    Since the estimators are sample averages of independent and identically distributed random vectors, parts (i) and (ii) follow from the multivariate central limit theorem.
    Indeed, the scalar central limit theorem applies to every fixed linear combination of the contributions; if that combination has variance zero, its centered value is zero almost surely.
    The Cram\'er--Wold argument therefore also covers singular covariance matrices.
    Part (iii) follows from the strong law of large numbers applied to the contributions and to their products.
\end{proofEnd}
Because the lower and upper estimators are computed from the same individuals, they are generally correlated, and $\Sigma_j$ and $\Sigma_{jk}$ keep this correlation.
The covariance matrix can be singular; for example, $\Sigma_{jk}$ is the zero matrix if the two models output the same predictions.
The limit is normal regardless of whether the value of $r$ attaining the infimum or supremum for each individual is unique, because the optimization over $r$ takes place inside the expectation, individual by individual.
Section~\ref{subsec:confidence-tests} uses this result to construct confidence regions and tests of robust strict dominance.

The estimator of the upper bound on the difference from the best competitor is a maximum of sample averages, so its limiting distribution is generally not normal.
Let
\[
    \mathcal K_j^\ast
    =
    \operatorname*{arg\,max}_{k\ne j}\overline{\operatorname{IBS}}_{jk}(\tau;P)
\]
be the set of competitors attaining the upper bound in Proposition~\ref{prop:best-bounds}.
This set is nonempty because the number of competitors is finite.

\begin{theoremEnd}{theorem}[Asymptotic distribution of the upper bound for the best competitor]\label{thm:best-asymptotics}
    Under Assumptions~\ref{assump:iid}, \ref{assump:survival}, and~\ref{assump:independent-training}, for each model $j$, the vector
    \[
        \sqrt n\bigl\{
        \widehat{\overline{\operatorname{IBS}}}_{jk}(\tau)
        -\overline{\operatorname{IBS}}_{jk}(\tau;P)
        \bigr\}_{k\ne j}
    \]
    converges in distribution to a mean-zero normal vector $Z_j=(Z_{jk})_{k\ne j}$ with
    \[
        \operatorname{Cov}(Z_{jk},Z_{jk'})
        =
        \operatorname{Cov}_P\{\overline d_{jk}(O),\overline d_{jk'}(O)\}.
    \]
    Moreover,
    \[
        \sqrt n\bigl\{\widehat{\overline M}_j(\tau)-\overline M_j(\tau;P)\bigr\}
        \xrightarrow{d}
        \max_{k\in\mathcal K_j^\ast}Z_{jk}.
    \]
\end{theoremEnd}
\begin{proofEnd}
    The vector of pairwise upper-bound estimators is a sample average of independent and identically distributed bounded vectors.
    The multivariate central limit theorem therefore gives the stated normal limit and covariance matrix, including when that matrix is singular.

    Any competitor outside $\mathcal K_j^\ast$ has a population upper bound strictly smaller than $\overline M_j(\tau;P)$.
    There are finitely many such competitors, and Proposition~\ref{prop:consistency}(ii) gives simultaneous almost sure convergence of all pairwise upper-bound estimators.
    Consequently, with probability one, no competitor outside $\mathcal K_j^\ast$ maximizes the estimated upper bounds once $n$ is sufficiently large.
    It follows that, eventually almost surely,
    \[
        \sqrt n\bigl\{\widehat{\overline M}_j(\tau)-\overline M_j(\tau;P)\bigr\}
        =
        \max_{k\in\mathcal K_j^\ast}
        \sqrt n\bigl\{
        \widehat{\overline{\operatorname{IBS}}}_{jk}(\tau)
        -\overline{\operatorname{IBS}}_{jk}(\tau;P)
        \bigr\}.
    \]
    The maximum over this fixed finite set is a continuous function, so the continuous mapping theorem proves the result.
\end{proofEnd}
If $\mathcal K_j^\ast$ contains several competitors, the limit is the maximum of several correlated normal variables and is not normal in general.
If $\mathcal K_j^\ast=\{k^\ast\}$ is a singleton, the limit reduces to the normal variable $Z_{jk^\ast}$.
Section~\ref{subsec:confidence-tests} constructs a test and a one-sided upper confidence bound from the pairwise upper bounds without directly approximating the distribution of their maximum.

\subsection{Confidence regions and tests}\label{subsec:confidence-tests}

An estimated identified set reflects two kinds of uncertainty: the width of the identified set, which remains because some event times are unobserved, and the sampling variability from estimating its bounds, and a confidence region adds the latter to the former.
For the IBS of one model and the difference between two models, we construct confidence regions that contain the entire closure of the identified set with asymptotic probability $1-\alpha$ \parencite{2007Chernozhukov_EstimationConfidenceRegions}.
A confidence interval designed to cover the true value of the target, rather than the entire identified set, with asymptotic probability at least $1-\alpha$ is generally narrower \parencite{2004Imbens_ConfidenceIntervalsPartially,2008Romano_InferenceIdentifiableParameters,2009Stoye_MoreConfidenceIntervals}.
We take the identified set itself as the object of inference, because it summarizes what the observed data reveal about the target without restricting the dependence between event and censoring times.
A region that covers the identified set is the sample counterpart of this object, whereas an interval for the true value is calibrated to cover a single unknown point within the set.

Through the normal approximation to the bound estimators, we construct confidence regions for the IBS of a single model and the IBS difference between two models by widening the estimated bounds by a common multiple of their standard errors.
Let $\theta$ denote either $\operatorname{IBS}_j(\tau;Q)$ or $\operatorname{IBS}_{jk}(\tau;Q)$, and let $\underline\theta(P)$ and $\overline\theta(P)$ be its sharp bounds, estimated by $\widehat{\underline\theta}$ and $\widehat{\overline\theta}$ as in Section~\ref{subsec:estimators}.
Let $\underline\sigma^2$ and $\overline\sigma^2$ be the variances of the corresponding individual contributions, $\rho$ their correlation when both are positive, and $\widehat{\underline\sigma}$, $\widehat{\overline\sigma}$, and $\widehat\rho$ the sample counterparts.
The region extends $c\,\widehat{\underline\sigma}/\sqrt n$ below the estimated lower bound and $c\,\widehat{\overline\sigma}/\sqrt n$ above the estimated upper bound, where the critical value $c$ is common to the two endpoints.
The joint asymptotic normality in Theorem~\ref{thm:joint-normality} determines how large $c$ must be.

The critical value is chosen so that the region covers both bounds at once.
The region contains the lower bound $\underline\theta(P)$ if and only if
\[
    \widehat{\underline\theta}-\frac{c\,\widehat{\underline\sigma}}{\sqrt n}\le\underline\theta(P)
    \iff
    \frac{\sqrt n(\widehat{\underline\theta}-\underline\theta(P))}{\widehat{\underline\sigma}}\le c,
\]
and it contains the upper bound $\overline\theta(P)$ if and only if
\[
    \widehat{\overline\theta}+\frac{c\,\widehat{\overline\sigma}}{\sqrt n}\ge\overline\theta(P)
    \iff
    -\frac{\sqrt n(\widehat{\overline\theta}-\overline\theta(P))}{\widehat{\overline\sigma}}\le c.
\]
By Theorem~\ref{thm:joint-normality}, its coverage probability is therefore approximately $\Pr(\max\{Z_1,-Z_2\}\le c)$, where $Z_1$ and $Z_2$ are the limits of $\sqrt n(\widehat{\underline\theta}-\underline\theta(P))/\underline\sigma$ and $\sqrt n(\widehat{\overline\theta}-\overline\theta(P))/\overline\sigma$, respectively, and have correlation $\rho$.

Accordingly, for $\rho\in[-1,1]$, let $(Z_1,Z_2)$ be a bivariate normal vector with standard normal margins and correlation $\rho$, and define the critical value
\[
    c_{1-\alpha}(\rho)
    =
    \inf\bigl\{c:\Pr(\max\{Z_1,-Z_2\}\le c)\ge1-\alpha\bigr\}.
\]
When $\widehat{\underline\sigma}>0$ and $\widehat{\overline\sigma}>0$, we define
\[
    \mathcal C_{1-\alpha}
    =
    \Bigl[
    \widehat{\underline\theta}-\frac{c_{1-\alpha}(\widehat\rho)\,\widehat{\underline\sigma}}{\sqrt n},\
    \widehat{\overline\theta}+\frac{c_{1-\alpha}(\widehat\rho)\,\widehat{\overline\sigma}}{\sqrt n}
    \Bigr],
\]
and otherwise we set $\mathcal C_{1-\alpha}=\mathbb R$.

\begin{theoremEnd}{proposition}[Confidence region for one or two models]\label{prop:cr-pair}
    Suppose that Assumptions~\ref{assump:iid}, \ref{assump:survival}, and~\ref{assump:independent-training} hold and that $\underline\sigma>0$ and $\overline\sigma>0$.
    Then
    \[
        \Pr\bigl\{[\underline\theta(P),\overline\theta(P)]\subseteq\mathcal C_{1-\alpha}\bigr\}\to1-\alpha
        \qquad\text{as }n\to\infty.
    \]
\end{theoremEnd}
\begin{proofEnd}
    We first show that the estimated standard deviations and correlation converge to their population counterparts.
    By Theorem~\ref{thm:joint-normality}(iii) and continuity of the square root, $\widehat{\underline\sigma}\to\underline\sigma>0$ and $\widehat{\overline\sigma}\to\overline\sigma>0$ almost surely.
    Since almost sure convergence implies convergence in probability, the probability that $\widehat{\underline\sigma}>0$ and $\widehat{\overline\sigma}>0$ tends to one, and the convention $\mathcal C_{1-\alpha}=\mathbb R$ does not affect the limit.
    Correlation is a continuous function of the covariance matrix when both marginal variances are positive, so $\widehat\rho\to\rho$ almost surely as well.

    Next, we reduce coverage to an event about the larger of the two standardized errors.
    On the event that $\widehat{\underline\sigma}>0$ and $\widehat{\overline\sigma}>0$, $[\underline\theta(P),\overline\theta(P)]\subseteq\mathcal C_{1-\alpha}$ holds if and only if
    \[
        \max\Bigl\{\frac{\sqrt n(\widehat{\underline\theta}-\underline\theta(P))}{\widehat{\underline\sigma}},\
        -\frac{\sqrt n(\widehat{\overline\theta}-\overline\theta(P))}{\widehat{\overline\sigma}}\Bigr\}
        \le c_{1-\alpha}(\widehat\rho).
    \]
    By Theorem~\ref{thm:joint-normality} and Slutsky's theorem, the vector inside the maximum converges in distribution to $(Z_1,-Z_2)$, where $(Z_1,Z_2)$ has standard normal margins and correlation $\rho$.
    For each $\rho\in[-1,1]$, let $F_\rho$ be the distribution function of $\max\{Z_1,-Z_2\}$.
    It therefore remains to show that $F_\rho$ is continuous and that $c_{1-\alpha}(\widehat\rho)\to c_{1-\alpha}(\rho)$ almost surely.

    For each $\rho\in[-1,1]$, $F_\rho$ is continuous on $\mathbb R$ and satisfies $F_\rho(c_{1-\alpha}(\rho)-\varepsilon)<1-\alpha<F_\rho(c_{1-\alpha}(\rho)+\varepsilon)$ for every $\varepsilon>0$.
    For $|\rho|<1$, both properties hold because $(Z_1,Z_2)$ has a positive density on $\mathbb R^2$.
    At $\rho=1$, we have $Z_2=Z_1$, so the maximum equals $|Z_1|$, whose distribution function is continuous on $\mathbb R$ and strictly increasing on $(0,\infty)$, which contains $c_{1-\alpha}(1)>0$.
    At $\rho=-1$, we have $Z_2=-Z_1$, so the maximum equals $Z_1$, which is standard normal.

    These two properties make $c_{1-\alpha}(\cdot)$ continuous on $[-1,1]$.
    To compare different values of $\rho$, including $\rho=\pm1$, write $(Z_1,Z_2)=(X,\rho X+\sqrt{1-\rho^2}Y)$ with independent standard normal $X$ and $Y$.
    For fixed $X$ and $Y$, $\max\{X,-\rho X-\sqrt{1-\rho^2}Y\}$ is continuous in $\rho$, so $\rho'\to\rho$ implies $F_{\rho'}\to F_\rho$ pointwise, by continuity of $F_\rho$.
    Evaluating at $c_{1-\alpha}(\rho)\pm\varepsilon$ shows that $F_{\rho'}(c_{1-\alpha}(\rho)-\varepsilon)<1-\alpha<F_{\rho'}(c_{1-\alpha}(\rho)+\varepsilon)$ for $\rho'$ close to $\rho$, and hence $|c_{1-\alpha}(\rho')-c_{1-\alpha}(\rho)|\le\varepsilon$.
    Since $\widehat\rho\to\rho$ almost surely, $c_{1-\alpha}(\widehat\rho)\to c_{1-\alpha}(\rho)$ almost surely, and the continuous mapping theorem and Slutsky's theorem give the limit $\Pr(\max\{Z_1,-Z_2\}\le c_{1-\alpha}(\rho))=1-\alpha$.
\end{proofEnd}
The critical value $c_{1-\alpha}(\widehat\rho)$ uses the correlation between the lower and upper estimators.
A simpler alternative is to replace it with the standard normal quantile $z_{1-\alpha/2}$, which by the Bonferroni inequality gives asymptotic coverage of at least $1-\alpha$ but can be conservative.
If a fixed benchmark value lies outside $\mathcal C_{1-\alpha}$, we conclude, at asymptotic level $\alpha$, that it does not belong to the closure of the identified set.

The tests below concern comparisons fixed before the evaluation sample is examined, as specified in Assumption~\ref{assump:prespecified}.

Whether model $j$ robustly strictly dominates model $k$ is tested by checking whether a one-sided upper confidence bound for the upper bound on the IBS difference is negative.
By Definition~\ref{def:robust-strict-dominance}, the hypotheses are
\[
    H_0:\overline{\operatorname{IBS}}_{jk}(\tau;P)\ge0
    \qquad\text{versus}\qquad
    H_1:\overline{\operatorname{IBS}}_{jk}(\tau;P)<0.
\]
Let $\overline\sigma_{jk}^2=\operatorname{Var}_P\{\overline d_{jk}(O)\}$, let $\widehat{\overline\sigma}_{jk}$ be the sample standard deviation of $\overline d_{jk}(O_1),\ldots,\overline d_{jk}(O_n)$, and let $z_{1-\alpha}$ be the $(1-\alpha)$ quantile of the standard normal distribution with $0<\alpha<1/2$.
Define the one-sided upper confidence bound
\[
    U_{jk}
    =
    \widehat{\overline{\operatorname{IBS}}}_{jk}(\tau)+z_{1-\alpha}\frac{\widehat{\overline\sigma}_{jk}}{\sqrt n},
\]
and reject $H_0$ if $U_{jk}<0$.

\begin{theoremEnd}{proposition}[Test of robust strict dominance]\label{prop:test-dominance}
    Under Assumptions~\ref{assump:iid}, \ref{assump:survival}, \ref{assump:independent-training}, and~\ref{assump:prespecified}, the following hold for each fixed $P$.
    \begin{enumerate}[label=(\roman*)]
        \item \emph{Level.}
              If $\overline{\operatorname{IBS}}_{jk}(\tau;P)\ge0$, then $\limsup_{n\to\infty}\Pr\{U_{jk}<0\}\le\alpha$.
        \item \emph{Consistency.}
              If $\overline{\operatorname{IBS}}_{jk}(\tau;P)<0$, then $\Pr\{U_{jk}<0\}\to1$.
        \item \emph{Upper confidence bound.}
              $\liminf_{n\to\infty}\Pr\{\overline{\operatorname{IBS}}_{jk}(\tau;P)\le U_{jk}\}\ge1-\alpha$.
    \end{enumerate}
\end{theoremEnd}
\begin{proofEnd}
    All three parts follow from the asymptotic normality of the estimated upper bound on the IBS difference.
    Assumption~\ref{assump:prespecified} fixes $j$, $k$, and $\tau$ before the evaluation sample is examined, so the following limits apply to a fixed pair and horizon.
    Write $\theta=\overline{\operatorname{IBS}}_{jk}(\tau;P)$, $\widehat\theta=\widehat{\overline{\operatorname{IBS}}}_{jk}(\tau)$, and $\sigma=\overline\sigma_{jk}$.
    By Theorem~\ref{thm:joint-normality}, $\sqrt n(\widehat\theta-\theta)$ converges in distribution to $\sigma Z$ with $Z$ standard normal, and $\widehat{\overline\sigma}_{jk}\to\sigma$ almost surely.

    \emph{Part (i).}
    We bound the rejection probability under $H_0$, treating the cases $\sigma>0$ and $\sigma=0$ separately.
    Under $\theta\ge0$, the event $U_{jk}<0$ implies $\sqrt n(\widehat\theta-\theta)+z_{1-\alpha}\widehat{\overline\sigma}_{jk}<-\sqrt n\,\theta\le0$.
    Suppose first that $\sigma>0$.
    The left-hand side converges in distribution to $\sigma(Z+z_{1-\alpha})$, whose distribution is continuous at zero, so $\limsup_n\Pr\{U_{jk}<0\}\le\Pr\{\sigma(Z+z_{1-\alpha})<0\}=\alpha$.
    Suppose next that $\sigma=0$.
    Since $\theta=\mathbb E_P\,\overline d_{jk}(O)$, this means $\mathbb E_P[\{\overline d_{jk}(O)-\theta\}^2]=\sigma^2=0$.
    A nonnegative random variable with mean zero equals zero almost surely, so $\overline d_{jk}(O)=\theta$ almost surely.
    Hence $\widehat\theta=\theta$ and $\widehat{\overline\sigma}_{jk}=0$ almost surely, and $U_{jk}=\theta\ge0$, so $H_0$ is never rejected.

    \emph{Part (ii).}
    Under $H_1$, the upper confidence bound converges to the negative value $\theta$, so the test rejects with probability tending to one.
    Proposition~\ref{prop:consistency}(ii) gives $\widehat\theta\to\theta$ almost surely, and Theorem~\ref{thm:joint-normality}(iii) gives $\widehat{\overline\sigma}_{jk}/\sqrt n\to0$ almost surely.
    Hence $U_{jk}\to\theta<0$ almost surely, so $\Pr\{U_{jk}<0\}\to1$.

    \emph{Part (iii).}
    The coverage of $U_{jk}$ follows from the same limit as in Part~(i).
    The event $\theta\le U_{jk}$ is equivalent to $\sqrt n(\widehat\theta-\theta)+z_{1-\alpha}\widehat{\overline\sigma}_{jk}\ge0$.
    If $\sigma>0$, its probability converges to $\Pr\{\sigma(Z+z_{1-\alpha})\ge0\}=1-\alpha$.
    If $\sigma=0$, then $U_{jk}=\theta$ almost surely, and the probability is one.
\end{proofEnd}
At the boundary $\overline{\operatorname{IBS}}_{jk}(\tau;P)=0$, the rejection probability converges to $\alpha$ if $\overline\sigma_{jk}>0$ and equals zero if $\overline\sigma_{jk}=0$, which occurs, for example, when the two models output the same predictions.
Testing whether model $k$ robustly strictly dominates model $j$ amounts to rejecting when $\widehat{\underline{\operatorname{IBS}}}_{jk}(\tau)-z_{1-\alpha}\widehat{\underline\sigma}_{jk}/\sqrt n>0$, since $\overline d_{kj}=-\underline d_{jk}$; here $\widehat{\underline\sigma}_{jk}$ is the sample standard deviation of $\underline d_{jk}(O_i)$.
Because $\widehat{\underline{\operatorname{IBS}}}_{jk}(\tau)\le\widehat{\overline{\operatorname{IBS}}}_{jk}(\tau)$ and $z_{1-\alpha}>0$, the tests in the two directions never reject simultaneously.
If $H_0$ is rejected, we conclude that model $j$ predicts better than model $k$ under every full-data distribution compatible with the observed data, whereas a failure to reject does not imply that model $k$ is at least as good as model $j$.

Whether model $j$ is robustly strictly best among the $K$ models is tested by an intersection--union test (IUT), which rejects only if the test in Proposition~\ref{prop:test-dominance} rejects against every competitor.
By Definition~\ref{def:robustly-strictly-best} and Proposition~\ref{prop:best-bounds}, $\overline M_j(\tau;P)=\max_{k\ne j}\overline{\operatorname{IBS}}_{jk}(\tau;P)$, so the hypotheses are
\[
    H_0:\bigcup_{k\ne j}\bigl\{\overline{\operatorname{IBS}}_{jk}(\tau;P)\ge0\bigr\}
    \qquad\text{versus}\qquad
    H_1:\bigcap_{k\ne j}\bigl\{\overline{\operatorname{IBS}}_{jk}(\tau;P)<0\bigr\}.
\]
The null hypothesis is a union of the pairwise null hypotheses, and the alternative is their intersection \parencite{1982Berger_MultiparameterHypothesisTesting,1996Berger_BioequivalenceTrialsIntersectionunion}.
We reject $H_0$ if
\[
    U_j^{\mathrm{IUT}}=\max_{k\ne j}U_{jk}<0,
\]
where each $U_{jk}$ is computed at level $\alpha$ without adjusting for the number of competitors.
The comparison is fixed in advance by Assumption~\ref{assump:prespecified}.

\begin{theoremEnd}{proposition}[Intersection--union test for a robustly strictly best model]\label{prop:test-best}
    Under Assumptions~\ref{assump:iid}, \ref{assump:survival}, \ref{assump:independent-training}, and~\ref{assump:prespecified}, the following hold for each fixed $P$.
    \begin{enumerate}[label=(\roman*)]
        \item \emph{Level.}
              If $\overline M_j(\tau;P)\ge0$, then $\limsup_{n\to\infty}\Pr\{U_j^{\mathrm{IUT}}<0\}\le\alpha$.
        \item \emph{Consistency.}
              If $\overline M_j(\tau;P)<0$, then $\Pr\{U_j^{\mathrm{IUT}}<0\}\to1$.
        \item \emph{Upper confidence bound.}
              $\liminf_{n\to\infty}\Pr\{\overline M_j(\tau;P)\le U_j^{\mathrm{IUT}}\}\ge1-\alpha$.
    \end{enumerate}
\end{theoremEnd}
\begin{proofEnd}
    Each part follows from Proposition~\ref{prop:test-dominance} applied to the pairwise statistics $U_{jk}$.
    By Assumption~\ref{assump:prespecified}, $j$, the set of compared models, and $\tau$ do not depend on the evaluation sample, so each $U_{jk}$ is the statistic in Proposition~\ref{prop:test-dominance} for a fixed pair and horizon.

    \emph{Part (i).}
    Under $H_0$, the IUT can reject only if the pairwise test against a competitor that satisfies the null hypothesis rejects.
    By Proposition~\ref{prop:best-bounds}, the condition $\overline M_j(\tau;P)\ge0$ is equivalent to $H_0$, so there is $k_0\ne j$ with $\overline{\operatorname{IBS}}_{jk_0}(\tau;P)\ge0$.
    The event $U_j^{\mathrm{IUT}}<0$ implies $U_{jk_0}<0$, so Proposition~\ref{prop:test-dominance}(i) gives $\limsup_n\Pr\{U_j^{\mathrm{IUT}}<0\}\le\limsup_n\Pr\{U_{jk_0}<0\}\le\alpha$.
    This argument does not depend on how many competitors satisfy the null hypothesis or on the correlation among the estimators.

    \emph{Part (ii).}
    Under $H_1$, every pairwise test rejects with probability tending to one, and so does the IUT.
    If $\overline M_j(\tau;P)<0$, then $\overline{\operatorname{IBS}}_{jk}(\tau;P)<0$ for every $k\ne j$.
    Proposition~\ref{prop:test-dominance}(ii) applied to the finitely many competitors therefore gives $\Pr\{U_{jk}<0\text{ for all }k\ne j\}\to1$, which is $\Pr\{U_j^{\mathrm{IUT}}<0\}\to1$.

    \emph{Part (iii).}
    The bound $U_j^{\mathrm{IUT}}$ covers $\overline M_j(\tau;P)$ whenever the pairwise bound for a maximizing competitor covers its own target.
    Fix $k^\ast\in\mathcal K_j^\ast$, the set of maximizing competitors defined before Theorem~\ref{thm:best-asymptotics}.
    Since $U_j^{\mathrm{IUT}}\ge U_{jk^\ast}$ and $\overline M_j(\tau;P)=\overline{\operatorname{IBS}}_{jk^\ast}(\tau;P)$, the event $\overline{\operatorname{IBS}}_{jk^\ast}(\tau;P)\le U_{jk^\ast}$ implies $\overline M_j(\tau;P)\le U_j^{\mathrm{IUT}}$, and Proposition~\ref{prop:test-dominance}(iii) gives the result.
\end{proofEnd}
\noindent Assumption~\ref{assump:prespecified} rules out choosing the candidate model or the evaluation horizon after examining the evaluation sample, which would create a multiplicity problem and could inflate the probability of a false rejection.
The IUT does not require estimating $\mathcal K_j^\ast$ or identifying a unique maximizing competitor.
It can be conservative, however, because when several competitors lie on the boundary of the null hypothesis, all of their pairwise tests must reject.
This test and its one-sided upper confidence bound are thus constructed from the asymptotic normality of the pairwise upper bounds alone.

The coverage and level guarantees above are asymptotic and hold for each fixed observed-data distribution.
They are not uniform over classes of observed-data distributions: for example, if a few individuals have large positive contributions and the rest have small negative ones, an evaluation sample that contains none of the former yields a negative estimate with a small sample standard deviation, and the tests can then reject falsely.
The phrase ``every full-data distribution compatible with the observed data'' in Definitions~\ref{def:robust-strict-dominance} and~\ref{def:robustly-strictly-best} addresses the identification problem for a single $P$.
A uniform level, by contrast, would address the sampling problem over many possible $P$, which is a different issue.
This paper does not consider tests with uniform guarantees.

\section{Simulation study}\label{sec:simulation}

This section examines how the sharp bounds in Section~\ref{sec:partial-identification} and the inference procedures in Section~\ref{sec:estimation-inference} behave in finite samples when the dependence between event and censoring times is controlled.
Section~\ref{subsec:sim-design} describes the data-generating process and the conditions shared by all experiments.
Sections~\ref{subsec:sim-single}--\ref{subsec:sim-multiple} then evaluate a single model, two models, and three models.

\subsection{Simulation design}\label{subsec:sim-design}

Because the simulations generate the latent event times, we can compare the full-data IBS numerically with the bounds that delimit the closure of the identified set.
As in Sections~\ref{subsec:framework} and~\ref{subsec:estimators}, every prediction rule is fixed independently of the evaluation sample, and no model is trained within the replications.
This separates the estimation error of the prediction models from the identification problem caused by censoring.
We approximate the population bounds and the full-data IBS using an independent sample of 500,000 individuals and use these approximations as reference values.
The censoring distribution and the shifts of the competing models in Sections~\ref{subsec:sim-two} and~\ref{subsec:sim-multiple} are calibrated on separate independent samples of 1,000,000 individuals.
On the reference sample, the censoring proportions and the calibrated bounds were within about 0.001 of their targets.
Hence the differences in the results below arise from dependent censoring and partial identification, not from how the models are trained.

We use the copula-based construction of \textcite{2023Gharari_CopulabasedDeepSurvivala}: a pair of uniform random variables is transformed by Weibull inverse survival functions to generate dependent event and censoring times.
Our design uses a Gaussian copula in place of their Clayton and Frank copulas, three covariates in place of ten, and a censoring distribution that does not depend on the covariates.
Our dependence grid also includes negative values of Kendall's tau.
The covariates $X=(X_1,X_2,X_3)$ are mutually independent, with $X_1,X_3\sim N(0,1)$ and $X_2\sim\operatorname{Bernoulli}(0.5)$.
The covariate $X_3$ affects neither the event time nor the censoring time; Section~\ref{subsec:sim-multiple} uses it only to construct competing models that err in opposite directions.
Given $X$, the event time follows the Weibull proportional hazards model
\[
    S_0(t\mid X)
    =
    \exp\bigl[-\lambda_T t^{\kappa_T}\exp\{\eta(X)\}\bigr],
    \qquad
    \eta(X)=0.8X_1+0.5X_2,
    \qquad
    \lambda_T=0.08,
    \qquad
    \kappa_T=1.4,
\]
and the censoring time has the survival function $S_C(c)=\exp(-\lambda_C c^{1.2})$, which does not depend on $X$.
We draw $(U_T,U_C)$ from a Gaussian copula and set $T=S_0^{-1}(U_T\mid X)$ and $C=S_C^{-1}(U_C)$.
We specify the strength of dependence by Kendall's tau $\tau_C$ of the copula, which determines the copula correlation parameter $\zeta=\sin(\pi\tau_C/2)$ \parencite{2006_IntroductionCopulas}.
The value $\tau_C=0$ gives $C\perp(T,X)$, since the censoring distribution does not depend on $X$.
Positive and negative values give dependent censoring in opposite directions.
Varying $\tau_C$ leaves the marginal distribution of the event time unchanged; the censoring scale $\lambda_C$ is recalibrated for each dependence level, as described below.

The width of the identified set depends directly on the proportion of individuals censored before the end of the evaluation period, $\mu_\tau=P(\Delta=0,Y<\tau)$, so we specify this proportion separately from the strength of dependence $\tau_C$.
The horizon $\tau$ is the 75th percentile of the marginal distribution of $T$.
All experiments use the same grid of conditions:
\[
    \tau_C\in\{-0.6,-0.3,0,0.3,0.6\},
    \quad
    \mu_\tau\in\{0.10,0.30,0.50\},
    \quad
    n\in\{250,500,1000\}.
\]
For each pair $(\tau_C,\mu_\tau)$, we calibrate $\lambda_C$ numerically to target the censoring proportion $\mu_\tau$.
Increasing $n$ from 250 to 1000 shows how closely the finite-sample behavior approaches the asymptotic theory for a fixed observed-data distribution in Section~\ref{subsec:asymptotics}.
We call each combination of $(\tau_C,\mu_\tau,n)$ a cell and replicate each cell $R=1{,}000$ times.
With this number of replications, an estimated proportion whose true value is 0.05 or 0.95 has a Monte Carlo standard error of about 0.0069.
The confidence level is $1-\alpha=0.95$, and the tests use the significance level $\alpha=0.05$.
For each of the 15 combinations of dependence and censoring proportion, we thus compare the behavior of the bounds and the inference procedures across sample sizes.

As a benchmark, we use IPCW estimation of the IBS and an IPCW test of the IBS difference, both of which assume independent censoring in the sense $C\perp(T,X)$.
The censoring survival function is estimated by the reverse Kaplan--Meier estimator without covariates \parencite{1958Kaplan_NonparametricEstimationIncomplete,1999Graf_AssessmentComparisonPrognostica}.
Standard errors are estimated by a nonparametric bootstrap that resamples individuals, with 500 resamples in every replication.
IPCW estimates a single value of the full-data IBS, whereas the proposed methods target the identified set.
The observed coverage and rejection proportions of IPCW are therefore not comparisons of the two approaches on a common criterion.
We report the IPCW results to show what conclusions an analysis that ignores dependent censoring would reach.

\subsection{Single model}\label{subsec:sim-single}

This subsection evaluates the identified set and the confidence region for the IBS when the prediction rule is the true conditional survival function, $\widehat S=S_0$, which removes misspecification of the prediction rule as a source of bias.
We examine three questions.
First, does the interval between the reference bounds contain the reference full-data IBS across the dependence settings?
Second, is the IPCW estimator biased under dependent censoring, and if so, does the bias shrink as the sample size grows?
Third, is the proportion of replications in which the confidence region in Proposition~\ref{prop:cr-pair} contains both reference bounds close to the nominal level 0.95?
The evaluation measures are the reference width of the identified set, the estimated bias of the IPCW estimator, this empirical coverage proportion, and the length of the confidence region.

In all 15 scenarios, the interval between the reference bounds contained the reference full-data IBS (Table~\ref{tab:sim-single-bounds}).
Because the event-time distribution and the prediction rule are the same in all scenarios, the full-data IBS was about 0.153 throughout, and only the observed-data distribution changed with $\tau_C$ and $\mu_\tau$.
The width of the identified set was determined mainly by the censoring proportion: it was 0.022--0.024 for $\mu_\tau=0.10$, 0.069--0.074 for $\mu_\tau=0.30$, and 0.119--0.139 for $\mu_\tau=0.50$, and it varied much less across $\tau_C$.

The IPCW estimator, by contrast, was biased under dependent censoring, and the bias did not shrink as the sample size grew.
Under independent censoring, $\tau_C=0$, the bias was at most 0.0002 in absolute value.
Under dependent censoring, its sign and size depended on both $\tau_C$ and $\mu_\tau$.
The bias was largest, $-0.0106$ at $n=1000$, for $\tau_C=-0.6$ and $\mu_\tau=0.30$, which is about 15\% of the width of the identified set in that scenario, and it stayed within 0.0022 in absolute value for $\tau_C=0.3$.
In every scenario, the estimated bias was almost the same for $n=250$, 500, and 1000.
This pattern is consistent with bias caused by violation of the independent-censoring assumption.

\begin{table}[tbp]
    \centering
    \small
    \caption{Population bounds on the IBS of the true survival function $S_0$ and bias of the IPCW estimator.
        Lower, Upper, and Width approximate the sharp bounds and their difference, and Full data approximates the full-data IBS; all four are computed from the reference sample of 500,000 individuals in each scenario.
        IPCW bias is the mean over 1,000 replications of the IPCW estimate minus the full-data IBS.}
    \label{tab:sim-single-bounds}
    \begin{tabular}{rrrrrrrrr}
\toprule
& & \multicolumn{4}{c}{Population reference} & \multicolumn{3}{c}{IPCW bias} \\
\cmidrule(lr){3-6}\cmidrule(lr){7-9}
$\tau_C$ & $\mu_\tau$ & Lower & Upper & Width & Full data & $n=250$ & $n=500$ & $n=1000$ \\
\midrule
$-$0.6 & 0.10 & 0.1428 & 0.1646 & 0.0218 & 0.1531 & $-$0.0052 & $-$0.0054 & $-$0.0052 \\
$-$0.3 & 0.10 & 0.1459 & 0.1681 & 0.0222 & 0.1533 & $-$0.0022 & $-$0.0024 & $-$0.0026 \\
0.0 & 0.10 & 0.1473 & 0.1702 & 0.0229 & 0.1531 & $-$0.0001 & 0.0000 & $-$0.0001 \\
0.3 & 0.10 & 0.1467 & 0.1699 & 0.0232 & 0.1530 & $-$0.0004 & $-$0.0009 & $-$0.0007 \\
0.6 & 0.10 & 0.1423 & 0.1660 & 0.0237 & 0.1528 & $-$0.0044 & $-$0.0048 & $-$0.0047 \\
\addlinespace
$-$0.6 & 0.30 & 0.1296 & 0.2010 & 0.0714 & 0.1531 & $-$0.0102 & $-$0.0107 & $-$0.0106 \\
$-$0.3 & 0.30 & 0.1336 & 0.2068 & 0.0732 & 0.1531 & $-$0.0036 & $-$0.0039 & $-$0.0040 \\
0.0 & 0.30 & 0.1352 & 0.2092 & 0.0740 & 0.1529 & 0.0001 & 0.0002 & 0.0001 \\
0.3 & 0.30 & 0.1344 & 0.2067 & 0.0724 & 0.1528 & 0.0002 & 0.0001 & 0.0001 \\
0.6 & 0.30 & 0.1297 & 0.1985 & 0.0688 & 0.1531 & $-$0.0063 & $-$0.0058 & $-$0.0057 \\
\addlinespace
$-$0.6 & 0.50 & 0.1214 & 0.2607 & 0.1393 & 0.1530 & $-$0.0070 & $-$0.0075 & $-$0.0072 \\
$-$0.3 & 0.50 & 0.1223 & 0.2604 & 0.1381 & 0.1533 & $-$0.0045 & $-$0.0045 & $-$0.0043 \\
0.0 & 0.50 & 0.1223 & 0.2577 & 0.1354 & 0.1532 & 0.0001 & 0.0000 & $-$0.0001 \\
0.3 & 0.50 & 0.1213 & 0.2503 & 0.1290 & 0.1529 & 0.0022 & 0.0015 & 0.0021 \\
0.6 & 0.50 & 0.1180 & 0.2371 & 0.1192 & 0.1530 & $-$0.0015 & $-$0.0015 & $-$0.0015 \\
\bottomrule
\end{tabular}

\end{table}

In every cell, $\mathcal C_{1-\alpha}$ contained both reference bounds in close to 95\% of the 1,000 replications, and its excess length over the identified set shrank as the sample size grew (Table~\ref{tab:sim-single-inference}).
These empirical coverage proportions ranged from 0.935 to 0.967 over the 45 cells, and their averages over the 15 scenarios were 0.951, 0.952, and 0.949 for $n=250$, 500, and 1000.
The mean length of the region exceeded the width of the identified set by 0.027--0.032 for $n=250$ and by 0.013--0.016 for $n=1000$.
As the sample size grows, the part of the length due to sampling variability shrinks, while the width of the identified set remains.

For reference, Table~\ref{tab:sim-single-inference} also reports how often the 95\% IPCW interval contained the full-data IBS.
Under independent censoring, this proportion was 0.931--0.962, close to the nominal level.
Under dependent censoring, it fell to as low as 0.117, for $\tau_C=-0.6$, $\mu_\tau=0.30$, and $n=1000$.
For $\tau_C=-0.6$, $-0.3$, and 0.6, it decreased from $n=250$ to $n=1000$ in every scenario, alongside the estimated bias that changed little over these sample sizes.
These proportions concern a single value rather than the identified set, so they do not compare the two approaches on a common criterion.
For these three values of $\tau_C$, increasing the sample size over the range examined did not improve the IPCW interval's coverage of the full-data IBS.

\begin{table}[tbp]
    \centering
    \small
    \caption{Coverage and length of the 95\% confidence regions for the IBS of the true survival function $S_0$.
        Coverage is the proportion of 1,000 replications in which $\mathcal C_{1-\alpha}$ contained both reference bounds in Table~\ref{tab:sim-single-bounds}, and Mean length is the mean length of $\mathcal C_{1-\alpha}$.
        IPCW coverage is the proportion in which the 95\% IPCW interval contained the full-data IBS.
        Column headings give the sample size $n$.}
    \label{tab:sim-single-inference}
    \begin{tabular}{rrrrrrrrrrr}
\toprule
& & \multicolumn{3}{c}{Coverage} & \multicolumn{3}{c}{Mean length} & \multicolumn{3}{c}{IPCW coverage} \\
\cmidrule(lr){3-5}\cmidrule(lr){6-8}\cmidrule(lr){9-11}
$\tau_C$ & $\mu_\tau$ & $250$ & $500$ & $1000$ & $250$ & $500$ & $1000$ & $250$ & $500$ & $1000$ \\
\midrule
$-$0.6 & 0.10 & 0.938 & 0.935 & 0.951 & 0.0489 & 0.0407 & 0.0353 & 0.837 & 0.756 & 0.644 \\
$-$0.3 & 0.10 & 0.942 & 0.936 & 0.947 & 0.0497 & 0.0418 & 0.0360 & 0.923 & 0.901 & 0.867 \\
0.0 & 0.10 & 0.945 & 0.947 & 0.941 & 0.0511 & 0.0425 & 0.0368 & 0.931 & 0.946 & 0.946 \\
0.3 & 0.10 & 0.947 & 0.946 & 0.953 & 0.0515 & 0.0431 & 0.0374 & 0.947 & 0.941 & 0.950 \\
0.6 & 0.10 & 0.942 & 0.949 & 0.945 & 0.0504 & 0.0427 & 0.0373 & 0.880 & 0.814 & 0.692 \\
\addlinespace
$-$0.6 & 0.30 & 0.959 & 0.955 & 0.955 & 0.1012 & 0.0926 & 0.0865 & 0.644 & 0.382 & 0.117 \\
$-$0.3 & 0.30 & 0.955 & 0.956 & 0.949 & 0.1038 & 0.0946 & 0.0885 & 0.909 & 0.864 & 0.783 \\
0.0 & 0.30 & 0.941 & 0.947 & 0.943 & 0.1046 & 0.0956 & 0.0892 & 0.938 & 0.945 & 0.956 \\
0.3 & 0.30 & 0.961 & 0.953 & 0.947 & 0.1029 & 0.0938 & 0.0876 & 0.951 & 0.954 & 0.936 \\
0.6 & 0.30 & 0.947 & 0.949 & 0.944 & 0.0974 & 0.0888 & 0.0830 & 0.847 & 0.754 & 0.647 \\
\addlinespace
$-$0.6 & 0.50 & 0.962 & 0.967 & 0.958 & 0.1716 & 0.1626 & 0.1557 & 0.843 & 0.726 & 0.582 \\
$-$0.3 & 0.50 & 0.959 & 0.949 & 0.952 & 0.1704 & 0.1607 & 0.1544 & 0.898 & 0.858 & 0.798 \\
0.0 & 0.50 & 0.953 & 0.963 & 0.962 & 0.1672 & 0.1580 & 0.1509 & 0.939 & 0.953 & 0.962 \\
0.3 & 0.50 & 0.953 & 0.959 & 0.944 & 0.1613 & 0.1516 & 0.1445 & 0.936 & 0.946 & 0.926 \\
0.6 & 0.50 & 0.962 & 0.964 & 0.948 & 0.1486 & 0.1407 & 0.1342 & 0.932 & 0.941 & 0.915 \\
\bottomrule
\end{tabular}

\end{table}

\subsection{Comparison of two models}\label{subsec:sim-two}

We compare two prediction models: model 1 is the true conditional survival function, $\widehat S_1=S_0$, and model 2 adds a positive shift $\delta$ to the linear predictor,
\[
    \widehat S_2(t\mid X)
    =
    \exp\bigl[-\lambda_T t^{\kappa_T}\exp\{\eta(X)+\delta\}\bigr],
    \qquad
    \delta>0.
\]
Model 2 predicts a lower survival probability than model 1 for every individual at every time.
Because model 1 is the true survival function, it has the smaller IBS under the true full-data distribution.
When $\overline{\operatorname{IBS}}_{12}(\tau;P)>0$, that is, in the interior of the null hypothesis $\overline{\operatorname{IBS}}_{12}(\tau;P)\ge0$ of the test in Proposition~\ref{prop:test-dominance}, some full-data distribution compatible with $P$ gives model~2 the smaller IBS, so the observed data alone cannot establish that model~1 is better.

We choose $\delta$ so that the upper bound $\overline{\operatorname{IBS}}_{12}(\tau;P)$, computed on a separate sample of 1,000,000 individuals (Section~\ref{subsec:sim-design}), takes each of eight prescribed values, called targets, which place the setting in the interior of the null hypothesis, on its boundary, or in the alternative.
As $\delta$ increases from zero, the upper bound rises from zero to a positive maximum and then decreases, becoming negative for large $\delta$.
The targets are half of this maximum for the interior, zero for the boundary, and $-0.0025$, $-0.005$, $-0.01$, $-0.02$, $-0.04$, and $-0.06$ for the alternative.
The eight targets with the same $(\tau_C,\mu_\tau,n)$ share the same evaluation samples, because changing $\delta$ changes model 2 but not the data.
This design varies the target position relative to the null hypothesis while keeping the observed-data distribution fixed.

For the IBS difference, the proportion of 1,000 replications in which the confidence region contained both reference bounds was also close to the nominal level 0.95 (Table~\ref{tab:sim-two-null}).
Over the 360 combinations of cell and target level, these empirical coverage proportions ranged from 0.936 to 0.969, and their averages over the 120 combinations for each sample size were 0.951, 0.952, and 0.953.
The mean length of $\mathcal C_{1-\alpha}$ decreased with $n$ toward the width of the identified set; for $\mu_\tau=0.50$, it fell from 0.360 at $n=250$ to 0.325 at $n=1000$, while the width averaged 0.290.
With heavy censoring, most of the length of the region therefore reflects the width of the identified set rather than sampling variability.

At the zero target, the test of robust strict dominance rejected in close to 5\% of the replications (Table~\ref{tab:sim-two-null}).
The rejection proportions ranged from 0.035 to 0.061 over the 45 cells, and their averages over the 15 scenarios were 0.051, 0.049, and 0.048 for $n=250$, 500, and 1000.
These proportions are consistent with the statement after Proposition~\ref{prop:test-dominance} that, at the boundary, the rejection probability converges to $\alpha$ when $\overline\sigma_{12}>0$.

At the positive targets in the interior of the null hypothesis, the test rejected less often than at the zero target, and its average rejection proportion decreased as the sample size grew (Table~\ref{tab:sim-two-null}).
The rejection proportions were at most 0.050, and their averages over the 15 scenarios were 0.014, 0.008, and 0.005 for $n=250$, 500, and 1000.
The highest proportions occurred for $\mu_\tau=0.10$, where the interior targets were only 0.0001 to 0.0013 and thus close to zero.
These results agree with Proposition~\ref{prop:test-dominance}(i), which bounds the asymptotic rejection probability by $\alpha$ whenever the upper bound is nonnegative.

\begin{table}[tbp]
    \centering
    \small
    \caption{Empirical coverage for the IBS difference and rejection proportions at the intended null targets.
        Coverage is the proportion of 1,000 replications in which $\mathcal C_{1-\alpha}$ contained both reference bounds on $\operatorname{IBS}_{12}(\tau;Q)$, averaged over all eight target levels.
        Boundary rejection and Interior rejection are the proportions of 1,000 replications in which the test rejected at $\alpha=0.05$ when the upper bound was calibrated toward zero and toward half of its largest attainable value, respectively.
        These labels denote calibration targets rather than verified population null conditions.
        Column headings give the sample size $n$.}
    \label{tab:sim-two-null}
    \begin{tabular}{rrrrrrrrrrr}
\toprule
& & \multicolumn{3}{c}{Coverage} & \multicolumn{3}{c}{Boundary rejection} & \multicolumn{3}{c}{Interior rejection} \\
\cmidrule(lr){3-5}\cmidrule(lr){6-8}\cmidrule(lr){9-11}
$\tau_C$ & $\mu_\tau$ & $250$ & $500$ & $1000$ & $250$ & $500$ & $1000$ & $250$ & $500$ & $1000$ \\
\midrule
$-$0.6 & 0.10 & 0.951 & 0.955 & 0.952 & 0.042 & 0.061 & 0.054 & 0.017 & 0.017 & 0.007 \\
$-$0.3 & 0.10 & 0.945 & 0.956 & 0.955 & 0.061 & 0.048 & 0.040 & 0.028 & 0.012 & 0.006 \\
0.0 & 0.10 & 0.953 & 0.946 & 0.953 & 0.054 & 0.056 & 0.045 & 0.025 & 0.028 & 0.014 \\
0.3 & 0.10 & 0.950 & 0.943 & 0.951 & 0.056 & 0.044 & 0.045 & 0.034 & 0.024 & 0.018 \\
0.6 & 0.10 & 0.940 & 0.964 & 0.938 & 0.061 & 0.044 & 0.057 & 0.050 & 0.027 & 0.031 \\
\addlinespace
$-$0.6 & 0.30 & 0.946 & 0.949 & 0.962 & 0.045 & 0.048 & 0.040 & 0.001 & 0.000 & 0.000 \\
$-$0.3 & 0.30 & 0.957 & 0.956 & 0.967 & 0.049 & 0.041 & 0.035 & 0.002 & 0.001 & 0.000 \\
0.0 & 0.30 & 0.948 & 0.950 & 0.946 & 0.049 & 0.049 & 0.057 & 0.009 & 0.002 & 0.000 \\
0.3 & 0.30 & 0.963 & 0.944 & 0.962 & 0.037 & 0.051 & 0.036 & 0.010 & 0.002 & 0.000 \\
0.6 & 0.30 & 0.951 & 0.952 & 0.941 & 0.050 & 0.052 & 0.047 & 0.025 & 0.010 & 0.003 \\
\addlinespace
$-$0.6 & 0.50 & 0.951 & 0.948 & 0.956 & 0.057 & 0.053 & 0.055 & 0.000 & 0.000 & 0.000 \\
$-$0.3 & 0.50 & 0.952 & 0.950 & 0.955 & 0.051 & 0.049 & 0.053 & 0.000 & 0.000 & 0.000 \\
0.0 & 0.50 & 0.948 & 0.960 & 0.949 & 0.046 & 0.050 & 0.057 & 0.001 & 0.000 & 0.000 \\
0.3 & 0.50 & 0.959 & 0.951 & 0.945 & 0.045 & 0.044 & 0.045 & 0.004 & 0.000 & 0.001 \\
0.6 & 0.50 & 0.954 & 0.950 & 0.959 & 0.056 & 0.048 & 0.052 & 0.009 & 0.001 & 0.000 \\
\bottomrule
\end{tabular}

\end{table}

\begin{figure}[tbp]
    \centering
    \includegraphics[width=\textwidth]{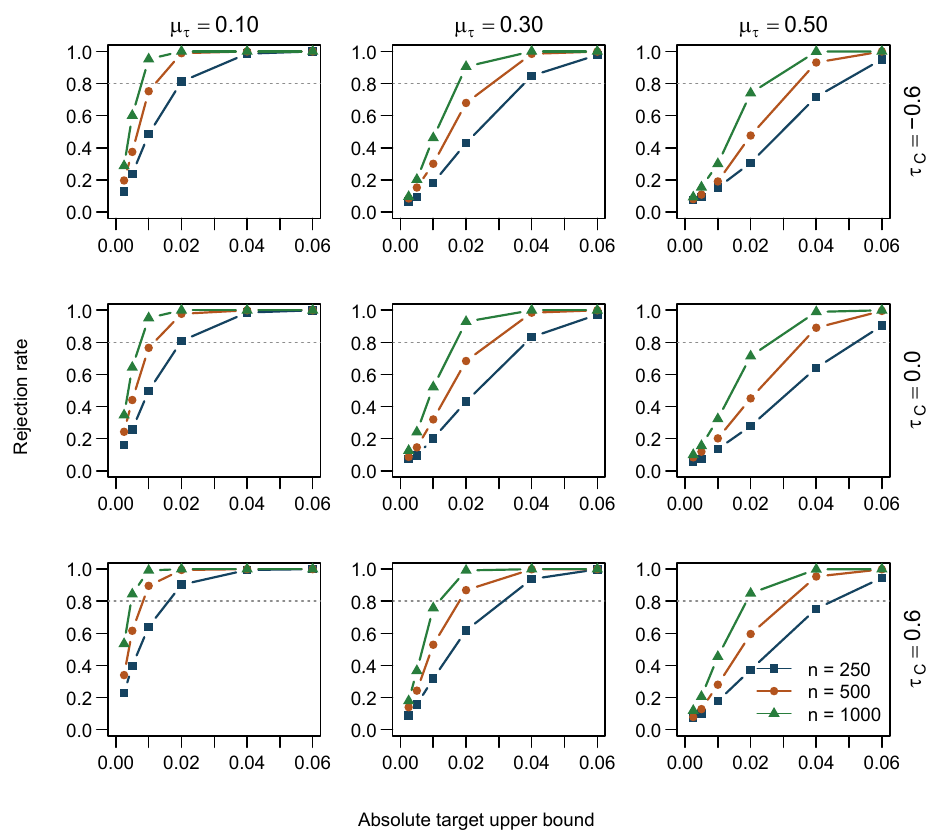}
    \caption{Power of the test of robust strict dominance at $\alpha=0.05$.
        The horizontal axis is the absolute value of the target upper bound $\overline{\operatorname{IBS}}_{12}(\tau;P)$, and the vertical axis is the proportion of 1,000 replications in which the test rejected.
        The rows correspond to the dependence levels $\tau_C=-0.6$, 0, and 0.6, the columns to the censoring proportion $\mu_\tau$, and the lines to the sample size $n$; the dotted line marks 0.8.}
    \label{fig:sim-two-power}
\end{figure}

The power of the test increased with the absolute value of the upper bound and with the sample size, and decreased with the censoring proportion (Figure~\ref{fig:sim-two-power}).
When the absolute value of the upper bound was 0.01 and $n=1000$, the power was 0.95 to 0.99 for $\mu_\tau=0.10$ but 0.30 to 0.46 for $\mu_\tau=0.50$ across the three dependence levels shown.
The power was also higher for $\tau_C=0.6$ than for $\tau_C=-0.6$ and 0, whose curves were similar; at the same absolute value and sample size with $\mu_\tau=0.30$, it was 0.46, 0.52, and 0.76 for $\tau_C=-0.6$, 0, and 0.6.
Heavier censoring before the horizon thus requires a larger difference in performance, or a larger sample, to establish robust strict dominance.

At the null targets, the IPCW test of the IBS difference rejected far more often than the test of robust strict dominance, but it addresses a different question.
This test rejects when the one-sided upper 95\% confidence bound for the IPCW estimate of the difference is negative.
Its rejection proportions averaged 0.83, 0.91, and 0.97 at the interior targets and 0.86, 0.95, and 0.99 at the zero target for $n=250$, 500, and 1000.
In this design, the full-data IBS difference was negative in every scenario and at every target level, so these rejections agree with the ranking under the true full-data distribution.
In the interior, however, some full-data distributions compatible with $P$ give model~2 the smaller IBS, and the IPCW test excludes them only through the untestable assumption of independent censoring.
Moreover, under dependent censoring the IPCW interval often failed to contain the full-data difference: averaged over the scenarios and target levels, its coverage at $n=1000$ was 0.95 for $\tau_C=0$ but 0.21 for $\tau_C=-0.6$ and 0.07 for $\tau_C=0.6$.
IPCW therefore ranks the models without reflecting the uncertainty created by dependent censoring, whereas the identified set and the test of robust strict dominance retain it.

\subsection{Comparison of multiple models}\label{subsec:sim-multiple}

We compare $K=3$ models in which model 1 is the true conditional survival function and models 2 and 3 are its competitors.
To show that model 1 is robustly strictly best, we must show that it robustly strictly dominates both model 2 and model 3, which requires two tests of the form in Proposition~\ref{prop:test-dominance}.
If the results of the two tests are reported as separate claims, the probability that at least one of them is false can exceed $\alpha$, as in multiple testing.
The IUT of Proposition~\ref{prop:test-best} instead combines the two tests into the single claim that model 1 is best, which it makes only if both tests reject, and its asymptotic level is at most $\alpha$ without any adjustment.
We examine its level and power with two kinds of competitors: (i) ordered competitors, for which one competitor alone determines the upper bound $\overline M_1(\tau;P)$, and (ii) symmetric competitors, for which the two pairwise upper bounds are equal.

With the ordered competitors (i), the IUT rejected at the zero target in close to 5\% of the replications.
Both competitors add a positive shift to the linear predictor as in Section~\ref{subsec:sim-two}, $\delta_2$ for model 2 and $\delta_3>\delta_2$ for model 3.
The shifts are chosen so that $\overline{\operatorname{IBS}}_{12}(\tau;P)$ takes the eight targets of Section~\ref{subsec:sim-two} and $\overline{\operatorname{IBS}}_{13}(\tau;P)$ takes values 0.005 smaller, so that $\overline M_1(\tau;P)=\overline{\operatorname{IBS}}_{12}(\tau;P)$ and model 3 is the weaker competitor.
At the zero target, the IUT rejected in proportions of 0.051, 0.049, and 0.048 on average for $n=250$, 500, and 1000, and from 0.035 to 0.060 in individual scenarios (Table~\ref{tab:sim-multiple-null}).
At least one of the two pairwise tests rejected in 0.16, 0.25, and 0.37 on average, and in up to 0.84 in individual scenarios.
These rejections come from the test against model 3, whose null hypothesis is false at this target, but model 1 is not robustly strictly best, so they cannot be read as evidence that it is.
The IUT is thus not always conservative: when one competitor alone determines the upper bound, its rejection proportion on the null boundary is close to $\alpha$.

The symmetric competitors (ii) shift the coefficient on $X_3$ by $\delta>0$ in opposite directions:
\[
    \widehat S_{2}(t\mid X)
    =
    \exp\bigl[-\lambda_T t^{\kappa_T}\exp\{\eta(X)+\delta X_3\}\bigr],
    \qquad
    \widehat S_{3}(t\mid X)
    =
    \exp\bigl[-\lambda_T t^{\kappa_T}\exp\{\eta(X)-\delta X_3\}\bigr].
\]
Model 2 predicts a lower survival probability than model 1 for individuals with $X_3>0$ and a higher one for those with $X_3<0$, and model 3 does the opposite.
Because $X_3$ is symmetric about zero and independent of $(X_1,X_2,T,C)$, the two pairwise upper bounds are equal, and we choose $\delta$ so that their common value $\overline M_1(\tau;P)$ takes the eight targets of Section~\ref{subsec:sim-two}.
At the zero target, both pairwise null hypotheses therefore lie on their boundaries, and the IUT rejects only if the two pairwise tests reject in the same sample.

With the symmetric competitors at the zero target, each pairwise test rejected in close to 5\% of the replications and at least one of them in close to 10\%, whereas the IUT almost never rejected (Table~\ref{tab:sim-multiple-null}).
Averaged over the 15 scenarios, the tests against models 2 and 3 rejected in proportions from 0.047 to 0.050, at least one of them rejected in 0.094, 0.096, and 0.097 for $n=250$, 500, and 1000, and the IUT rejected in 0.0006, 0.0004, and 0.0005.
In individual scenarios, the separate tests rejected in proportions from 0.032 to 0.066, at least one of them in 0.073 to 0.125, and the IUT in at most 0.003.
Because both pairwise null hypotheses hold, reporting the two results separately would thus make at least one false claim with probability about $2\alpha$.
At the positive targets in the interior of the null hypothesis, the IUT never rejected.
These proportions of the IUT are consistent with Proposition~\ref{prop:test-best}(i), which only bounds its asymptotic rejection probability by $\alpha$ from above.

\begin{table}[tbp]
    \centering
    \small
    \setlength{\tabcolsep}{4pt}
    \caption{Rejection proportions at the zero target, where the shifts were chosen so that $\overline M_1(\tau;P)=0$.
        Each entry is the proportion of 1,000 replications in which the test rejected at $\alpha=0.05$.
        For each kind of competitors, Either is the rejection of at least one of the tests of robust strict dominance of model 1 over models 2 and 3.
        Column headings give the sample size $n$.}
    \label{tab:sim-multiple-null}
    \begin{tabular}{rrrrrrrrrrrrrr}
\toprule
& & \multicolumn{6}{c}{(i) Ordered} & \multicolumn{6}{c}{(ii) Symmetric} \\
\cmidrule(lr){3-8}\cmidrule(lr){9-14}
& & \multicolumn{3}{c}{Either} & \multicolumn{3}{c}{IUT} & \multicolumn{3}{c}{Either} & \multicolumn{3}{c}{IUT} \\
\cmidrule(lr){3-5}\cmidrule(lr){6-8}\cmidrule(lr){9-11}\cmidrule(lr){12-14}
$\tau_C$ & $\mu_\tau$ & $250$ & $500$ & $1000$ & $250$ & $500$ & $1000$ & $250$ & $500$ & $1000$ & $250$ & $500$ & $1000$ \\
\midrule
$-$0.6 & 0.10 & 0.238 & 0.374 & 0.601 & 0.042 & 0.061 & 0.054 & 0.105 & 0.116 & 0.080 & 0.000 & 0.000 & 0.000 \\
$-$0.3 & 0.10 & 0.242 & 0.368 & 0.591 & 0.061 & 0.048 & 0.040 & 0.103 & 0.125 & 0.108 & 0.000 & 0.000 & 0.000 \\
0.0 & 0.10 & 0.258 & 0.441 & 0.644 & 0.054 & 0.056 & 0.045 & 0.096 & 0.085 & 0.091 & 0.000 & 0.000 & 0.000 \\
0.3 & 0.10 & 0.299 & 0.486 & 0.717 & 0.056 & 0.044 & 0.045 & 0.091 & 0.074 & 0.109 & 0.000 & 0.000 & 0.000 \\
0.6 & 0.10 & 0.398 & 0.616 & 0.844 & 0.061 & 0.044 & 0.057 & 0.120 & 0.097 & 0.093 & 0.000 & 0.000 & 0.000 \\
\addlinespace
$-$0.6 & 0.30 & 0.093 & 0.152 & 0.202 & 0.045 & 0.048 & 0.040 & 0.099 & 0.103 & 0.110 & 0.000 & 0.000 & 0.000 \\
$-$0.3 & 0.30 & 0.109 & 0.129 & 0.215 & 0.049 & 0.041 & 0.035 & 0.094 & 0.084 & 0.097 & 0.000 & 0.000 & 0.001 \\
0.0 & 0.30 & 0.095 & 0.146 & 0.241 & 0.049 & 0.049 & 0.057 & 0.073 & 0.095 & 0.097 & 0.001 & 0.000 & 0.000 \\
0.3 & 0.30 & 0.112 & 0.199 & 0.272 & 0.037 & 0.051 & 0.036 & 0.096 & 0.088 & 0.106 & 0.001 & 0.001 & 0.000 \\
0.6 & 0.30 & 0.157 & 0.243 & 0.367 & 0.050 & 0.052 & 0.047 & 0.085 & 0.102 & 0.089 & 0.000 & 0.000 & 0.001 \\
\addlinespace
$-$0.6 & 0.50 & 0.094 & 0.108 & 0.154 & 0.057 & 0.053 & 0.055 & 0.085 & 0.097 & 0.092 & 0.001 & 0.002 & 0.003 \\
$-$0.3 & 0.50 & 0.088 & 0.115 & 0.144 & 0.051 & 0.049 & 0.053 & 0.089 & 0.102 & 0.103 & 0.003 & 0.000 & 0.002 \\
0.0 & 0.50 & 0.072 & 0.119 & 0.156 & 0.046 & 0.050 & 0.057 & 0.087 & 0.101 & 0.104 & 0.001 & 0.001 & 0.000 \\
0.3 & 0.50 & 0.074 & 0.107 & 0.138 & 0.045 & 0.044 & 0.045 & 0.084 & 0.080 & 0.081 & 0.002 & 0.001 & 0.001 \\
0.6 & 0.50 & 0.102 & 0.128 & 0.206 & 0.056 & 0.048 & 0.052 & 0.099 & 0.089 & 0.091 & 0.000 & 0.001 & 0.000 \\
\bottomrule
\end{tabular}

\end{table}

The low rejection proportions of the IUT can be explained by a negative correlation between the upper-bound contributions against the two competitors.
An individual with $X_3>0$ receives a lower prediction from model 2 and a higher one from model 3, so its contributions $\overline d_{12}(O)$ and $\overline d_{13}(O)$ tend to move in opposite directions.
The test against model $k$ rejects when $U_{1k}<0$, that is, when $\widehat{\overline{\operatorname{IBS}}}_{1k}(\tau)<-z_{1-\alpha}\widehat{\overline\sigma}_{1k}/\sqrt n$.
Both estimates are sample averages over the same individuals, so they inherit the correlation of $\overline d_{12}(O)$ and $\overline d_{13}(O)$: a sample whose individuals happen to favor model 1 over model 2 tends to favor model 3 over model 1, and vice versa.
The IUT, however, rejects only if the inequality above holds for both $k=2$ and $k=3$ in the same sample, that is, only if the sample favors model 1 over both competitors at once.
The negative correlation makes such samples rare, so the IUT rarely rejects.

The observed rejection proportions of the IUT agreed with those implied by the correlations in the reference samples.
The correlation between the two contributions averaged $-0.78$, $-0.36$, and $-0.10$ for $\mu_\tau=0.10$, 0.30, and 0.50, and the IUT rejection proportion at the zero target rose from 0.000 to 0.0003 and 0.0012.
To compute the implied proportions, we approximate the two standardized estimates $\sqrt n\,\widehat{\overline{\operatorname{IBS}}}_{1k}(\tau)/\widehat{\overline\sigma}_{1k}$, $k=2,3$, by standard normal variables with these correlations, which is their limit when both upper bounds are zero.
The IUT rejects when both are below $-z_{1-\alpha}$, and this probability is about $2\times10^{-8}$, 0.0003, and 0.0016 for $\mu_\tau=0.10$, 0.30, and 0.50, close to the observed proportions.
This probability depends on the correlation: it would be $\alpha=0.05$ if the two estimates were perfectly positively correlated and $\alpha^2=0.0025$ if they were independent, and it is smaller still when they are negatively correlated.
With these competitors, the rejection proportions of the IUT on the null boundary were therefore far below the nominal level $\alpha$.

\begin{figure}[tbp]
    \centering
    \includegraphics[width=\textwidth]{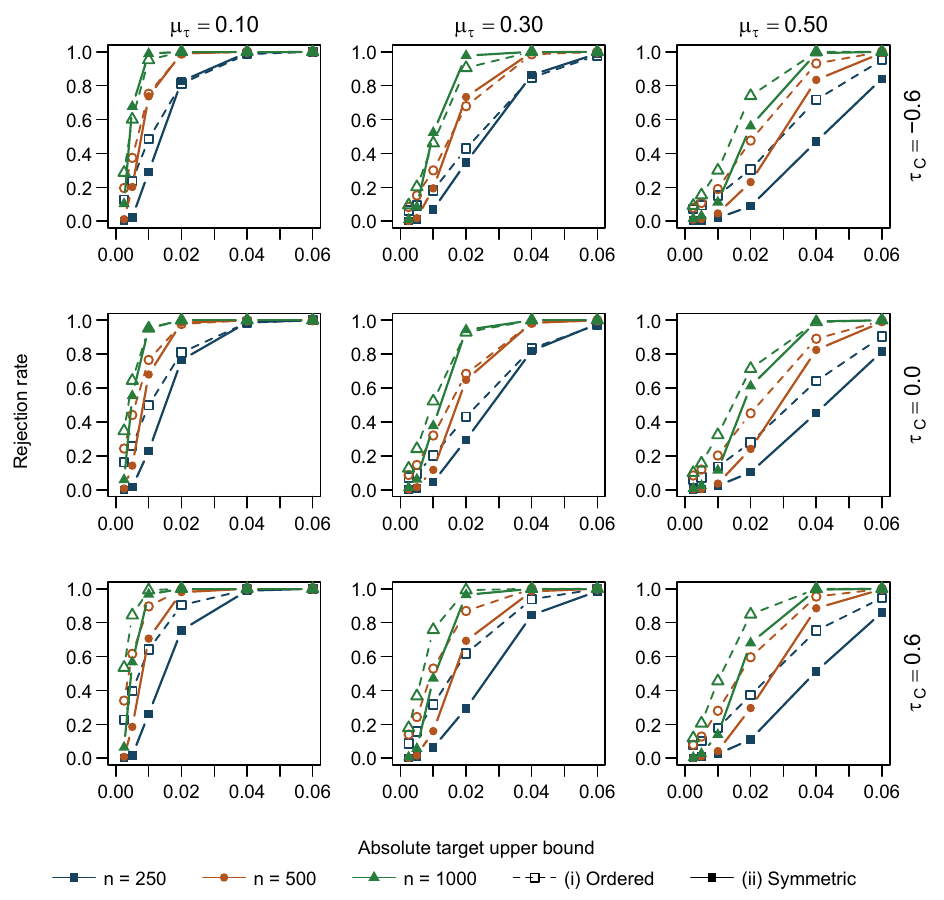}
    \caption{Power of the IUT at $\alpha=0.05$ with (i) the ordered and (ii) the symmetric competitors.
        The horizontal axis is the absolute value of the target upper bound $\overline M_1(\tau;P)$, and the vertical axis is the proportion of 1,000 replications in which the IUT rejected.
        The rows correspond to the dependence levels $\tau_C=-0.6$, 0, and 0.6, the columns to the censoring proportion $\mu_\tau$, and the colors to the sample size $n$.
        Dashed lines with open markers show the ordered competitors, and solid lines with filled markers show the symmetric competitors.}
    \label{fig:sim-multiple-power}
\end{figure}

At the same value of the upper bound, the IUT had lower power with the symmetric competitors (ii) than with the ordered competitors (i), and the difference was largest when the absolute value of the upper bound was small (Figure~\ref{fig:sim-multiple-power}).
Averaged over the 15 scenarios at $n=1000$, the power with the ordered and the symmetric competitors was 0.20 and 0.03 when the absolute value of the upper bound was 0.0025, 0.62 and 0.51 when it was 0.01, and 0.90 and 0.86 when it was 0.02, and both were at least 0.997 when it was 0.04.
The difference shrank as $n$ and the absolute value grew, because every pairwise test then rejects with probability tending to one, as Proposition~\ref{prop:test-best}(ii) states.
Heavier censoring lowered both curves, as in Section~\ref{subsec:sim-two}.
To show that a model is robustly strictly best when several competitors are equally close to it, a larger difference in performance or a larger sample is thus needed than when one competitor stands out.

\section{Real data applications}\label{sec:real-data}

This section applies the bounds in Section~\ref{sec:partial-identification} and the inference procedures in Section~\ref{sec:estimation-inference} to three real data sets and shows how the conclusions differ from those of IPCW-based evaluation, which assumes independent censoring.
Section~\ref{subsec:real-data-sets} introduces data on unemployment duration, churn from a music streaming service, and mortgage prepayment, which differ widely in the proportion censored within the evaluation period, and Section~\ref{subsec:real-procedure} describes the procedure common to the three data sets.
Section~\ref{subsec:real-results} reports the results for each data set and compares them across the three data sets.

\subsection{Data sets}\label{subsec:real-data-sets}

We use three data sets that differ widely in the proportion censored before the horizon and in sample size (Table~\ref{tab:real-data-summary}).
They record unemployment duration \parencite[UnempDur;][]{1996McCall_UnemploymentInsuranceRules}, churn from a music streaming service \parencite[KKBOX;][]{2017Kaggle_WSDMKKBoxsChurn,2024Kvamme_PycoxSurvivalAnalysis}, and mortgage prepayment in the Single Family Loan-Level Dataset of Freddie Mac \parencite[SFLLD;][]{2026FreddieMac_SingleFamilyLoanLevel}.
They contain 3,343, 50,000, and 36,115 individuals, of whom 56.3\%, 22.4\%, and 0.4\% are censored before the horizon, respectively.
These differences allow us to compare how the conclusions supported without restricting the dependence between event and censoring times change with the amount of censoring and the sample size.

UnempDur consists of the unemployment spells of 3,343 individuals, recorded in two-week units.
The event is reemployment in a full-time job, and spells ending in a part-time job, in a job of unknown type, or in continued unemployment are censored.
With the horizon $\tau=13$, or 26 weeks, 56.3\% of the individuals are censored before the horizon, the largest proportion among the three data sets.
The censoring may depend on the time to full-time reemployment even given the covariates, because individuals who accepted a part-time job may have had worse prospects of finding a full-time job.

KKBOX consists of the transaction records of a music streaming service released for a Kaggle competition, and the event is churn, defined as in the Python library pycox.
We draw 50,000 members from the first subscription spells of members registered on or after 1 January 2015, a restriction intended to avoid left truncation.
All spells are administratively censored on 29 January 2017 as in pycox, and with $\tau=365$ days, 22.4\% of the members are censored before the horizon.
Because the censoring time is thus determined by the start date of the subscription, independent censoring is more plausible here than in the other two data sets, but the bounds do not assume it.

SFLLD records the monthly performance of mortgages, and we use the 36,115 loans in its 2014 sample file whose first payment was due in 2014 and whose records start at loan age zero.
The event is voluntary prepayment, and loans that end in a credit-related disposition, a whole loan sale, a reperforming loan securitization, or a confirmed underwriting or major servicing defect before a credit event are censored.
With $\tau=60$ months, which ends before March 2020, only 0.4\% of the loans are censored before the horizon, the smallest proportion among the three data sets.
The censoring may depend on the time to prepayment even given the covariates, because delinquency may restrict access to refinancing, so loans ending in a credit-related disposition may have been prepaid later than other loans had they continued; this is our interpretation.

\begin{table}[tbp]
    \centering
    \small
    \caption{The three data sets.
        The horizon $\tau$ is in the recorded time unit of each data set, two-week units for UnempDur.
        Censored is the percentage of individuals censored before the horizon, with $\Delta=0$ and $Y<\tau$, and Columns is the number of covariate columns after encoding.}
    \label{tab:real-data-summary}
    \begin{tabular}{@{}lllrrr@{}}
\toprule
Data & Event & Horizon $\tau$ & $n$ & Censored (\%) & Columns \\
\midrule
UnempDur & Full-time job & 13 (26 weeks) & 3,343 & 56.3 & 6 \\
KKBOX & Churn & 365 days & 50,000 & 22.4 & 44 \\
SFLLD & Prepayment & 60 months & 36,115 & 0.4 & 28 \\
\bottomrule
\end{tabular}

\end{table}

\subsection{Analysis procedure}\label{subsec:real-procedure}

We split each data set once into training, validation, and test samples in proportions of 70\%, 10\%, and 20\%, stratified by the event indicator, and use the test sample only for evaluation.
The test sample is not used for fitting the models, selecting their hyperparameters, or choosing the model to be evaluated.
Continuous covariates are standardized, categorical covariates are encoded as indicator variables, and missing values are imputed, all using quantities computed from the training sample only.
The test sample serves as the evaluation sample of Section~\ref{sec:estimation-inference}, and the fitted models are independent of it, as Assumption~\ref{assump:independent-training} requires.

The three candidate models are fitted with hyperparameters selected on the validation sample.
They are the Cox proportional hazards model (CoxPH) \parencite{1972Cox_RegressionModelsLifeTables}, which has no hyperparameter to select, the random survival forest (RSF) \parencite{2008Ishwaran_RandomSurvivalForests}, and DeepSurv \parencite{2018Katzman_DeepSurvPersonalized}.
RSF uses 100 trees with a maximum depth of 3, and its grid varies the fraction of covariate columns tried at each split over $\{1/3,2/3\}$ and the minimum node size over $\{3,10\}$.
DeepSurv has two hidden layers of equal width with ReLU activations and dropout probability 0.1 and is trained by Adam with learning rate 0.01 and weight decay $10^{-4}$, and its grid varies the width over $\{16,64\}$ and the number of epochs over $\{25,50,100\}$.
For each model, we select the setting with the smallest IPCW estimate of the IBS on the validation sample and refit the model on the combined training and validation samples.
Before looking at the test sample, we fix the model with the smallest such estimate as the evaluated model of Assumption~\ref{assump:prespecified}, with the other two models as its competitors.
This rule separates model selection from evaluation, so the asymptotic level of the intersection--union test in Proposition~\ref{prop:test-best} is not invalidated by reselecting the model on the test sample.

On the test sample, we compute the estimators, confidence regions, and tests of Section~\ref{sec:estimation-inference} for single models, pairs of models, and the three models together.
For each model, we compute the estimated bounds on its IBS and the confidence region for the identified set in Proposition~\ref{prop:cr-pair}.
For each pair of models, we compute the estimated bounds on the IBS difference, the confidence region in the same proposition, and the tests of robust strict dominance in both directions in Proposition~\ref{prop:test-dominance}.
For the three models together, we estimate the upper bound $\overline M_j(\tau;P)$ on the difference of each model from its best competitor and, for the prespecified model, compute the intersection--union test in Proposition~\ref{prop:test-best}.
All tests use the significance level $\alpha=0.05$, and the confidence regions and tests rely on the asymptotic guarantees for a fixed observed-data distribution in Section~\ref{subsec:confidence-tests}.

We compare these results with IPCW, which estimates a single value of the full-data IBS under independent censoring.
The IPCW estimates are weighted by the reverse Kaplan--Meier estimator without covariates, and their 95\% confidence intervals are computed from 1,000 bootstrap resamples of individuals.
As in Section~\ref{subsec:sim-design}, this benchmark assumes $C\perp(T,X)$.
The IPCW confidence intervals concern a single value of the full-data IBS, whereas the confidence regions concern the identified set, and we keep this distinction when interpreting the results.

\subsection{Results}\label{subsec:real-results}

The estimated identified set for the IBS of each model was narrower in data sets with a smaller proportion censored before the horizon (panel~(a) of Tables~\ref{tab:real-results-unempdur}--\ref{tab:real-results-sflld}).
The widths were 0.138 to 0.150 for UnempDur, 0.059 to 0.071 for KKBOX, and 0.0011 to 0.0014 for SFLLD.
For SFLLD, the estimated range of IBS values compatible with the observed data thus had width about 0.001 for each model.
Only individuals censored before the horizon contribute to this width, but their contributions also depend on the predicted survival curves (Section~\ref{subsec:single-model}).

In all three data sets, the IPCW estimates lay within the estimated identified sets and particularly close to their lower bounds, at most 10\% of the width above them for UnempDur and at most 20\% for the other two (panel~(a) of Tables~\ref{tab:real-results-unempdur}--\ref{tab:real-results-sflld}).
For UnempDur and KKBOX, the estimated upper bounds exceeded the IPCW estimates by 0.05 to 0.14, and for UnempDur they were nearly twice the IPCW estimates.
According to these estimates, the IBS of each model in these two data sets could be considerably larger than the IPCW estimate under some full-data distributions compatible with the observed data.
The IPCW estimates alone therefore do not show how large the IBS could be without the assumption of independent censoring.

The estimated identified sets for the IBS difference between two models excluded zero for all three pairs in SFLLD but contained zero for all three pairs in UnempDur and KKBOX (panel~(b) of Tables~\ref{tab:real-results-unempdur}--\ref{tab:real-results-sflld}).
That the estimated identified sets contain zero does not establish that the population identified sets do.
In all data sets, the bounds on the differences were 8\% to 20\% as wide as those obtained by combining the separate bounds for the two models, because each individual's event time is shared by both models (Proposition~\ref{prop:pair-bounds}).
Only in SFLLD, therefore, did the estimates indicate the sign of every difference without restricting the dependence between event and censoring times.

The tests of robust strict dominance supported the IPCW ranking in SFLLD but rejected for no pair in UnempDur and KKBOX (panel~(b) of Tables~\ref{tab:real-results-unempdur}--\ref{tab:real-results-sflld}).
In SFLLD, the tests, each at the 5\% significance level without adjustment for multiple testing, supported DeepSurv over both CoxPH and RSF, and CoxPH over RSF.
In UnempDur, IPCW distinguished only CoxPH and RSF, and only barely, with a lower confidence limit of about $5\times10^{-6}$ that is rounded to 0.0000 in Table~\ref{tab:real-results-unempdur}.
In KKBOX, by contrast, the IPCW confidence intervals excluded zero for all three pairs.
The comparisons supported by these tests without restricting the dependence between event and censoring times therefore differed across the data sets.

\begin{table}[tbp]
    \centering
    \footnotesize
    \caption{Results for UnempDur on the test sample ($n=669$).
        (a) Estimated bounds on the IBS, the 95\% confidence region (CR) for the identified set in Proposition~\ref{prop:cr-pair}, and the IPCW estimate with its 95\% bootstrap confidence interval (CI).
        (b) For each pair $(j,k)$, estimated bounds on $\operatorname{IBS}_{jk}(\tau;Q)=\operatorname{IBS}_j(\tau;Q)-\operatorname{IBS}_k(\tau;Q)$, the 95\% CR, the one-sided upper confidence bounds of the tests of whether model~$j$ robustly strictly dominates model~$k$ ($U_{jk}$) and the reverse ($U_{kj}$) in Proposition~\ref{prop:test-dominance}, and the IPCW estimate of the difference with its 95\% CI.
        (c) For each model $j$, the estimate of the upper bound $\overline M_j(\tau;P)$ on the difference $M_j(\tau;Q)$ from the best competitor in Proposition~\ref{prop:best-bounds}, and $U_j^{\mathrm{IUT}}$ of the intersection--union test in Proposition~\ref{prop:test-best} for the prespecified model, marked with an asterisk.
        Each test rejects at $\alpha=0.05$ if its upper confidence bound is negative.}
    \label{tab:real-results-unempdur}
    \setlength{\tabcolsep}{3.5pt}
    \begin{tabular}{@{}lrrcrrrc@{}}
\toprule
\multicolumn{8}{l}{\textit{(a) IBS of each model}} \\
\addlinespace[2pt]
Model & Lower & Upper & 95\% CR & & & IPCW & 95\% CI \\
\midrule
CoxPH & 0.145 & 0.287 & [0.135, 0.299] &  &  & 0.158 & [0.145, 0.171] \\
RSF & 0.145 & 0.283 & [0.135, 0.295] &  &  & 0.154 & [0.141, 0.167] \\
DeepSurv & 0.139 & 0.290 & [0.129, 0.302] &  &  & 0.154 & [0.141, 0.168] \\
\midrule
\multicolumn{8}{l}{\textit{(b) IBS difference between two models}} \\
\addlinespace[2pt]
Pair & Lower & Upper & 95\% CR & $U_{jk}$ & $U_{kj}$ & IPCW & 95\% CI \\
\midrule
CoxPH $-$ RSF & $-$0.0144 & 0.0204 & [$-$0.0174, 0.0234] & 0.0229 & 0.0169 & 0.0036 & [0.0000, 0.0072] \\
CoxPH $-$ DeepSurv & $-$0.0161 & 0.0200 & [$-$0.0196, 0.0228] & 0.0224 & 0.0190 & 0.0033 & [$-$0.0003, 0.0070] \\
RSF $-$ DeepSurv & $-$0.0158 & 0.0141 & [$-$0.0187, 0.0163] & 0.0159 & 0.0182 & $-$0.0003 & [$-$0.0033, 0.0028] \\
\midrule
\multicolumn{8}{l}{\textit{(c) Difference from the best competitor}} \\
\addlinespace[2pt]
Model & & Upper & & $U_j^{\mathrm{IUT}}$ & & & \\
\midrule
CoxPH &  & 0.0204 &  &  &  &  &  \\
RSF &  & 0.0144 &  &  &  &  &  \\
DeepSurv$^\ast$ &  & 0.0161 &  & 0.0190 &  &  &  \\
\bottomrule
\end{tabular}

\end{table}

\begin{table}[tbp]
    \centering
    \footnotesize
    \caption{Results for KKBOX on the test sample ($n=10{,}002$).
        The panels and columns are as in Table~\ref{tab:real-results-unempdur}.}
    \label{tab:real-results-kkbox}
    \setlength{\tabcolsep}{3.5pt}
    \begin{tabular}{@{}lrrcrrrc@{}}
\toprule
\multicolumn{8}{l}{\textit{(a) IBS of each model}} \\
\addlinespace[2pt]
Model & Lower & Upper & 95\% CR & & & IPCW & 95\% CI \\
\midrule
CoxPH & 0.098 & 0.169 & [0.095, 0.173] &  &  & 0.110 & [0.106, 0.113] \\
RSF & 0.107 & 0.166 & [0.104, 0.170] &  &  & 0.118 & [0.115, 0.121] \\
DeepSurv & 0.095 & 0.165 & [0.092, 0.169] &  &  & 0.107 & [0.103, 0.110] \\
\midrule
\multicolumn{8}{l}{\textit{(b) IBS difference between two models}} \\
\addlinespace[2pt]
Pair & Lower & Upper & 95\% CR & $U_{jk}$ & $U_{kj}$ & IPCW & 95\% CI \\
\midrule
CoxPH $-$ RSF & $-$0.0126 & 0.0053 & [$-$0.0140, 0.0069] & 0.0066 & 0.0137 & $-$0.0081 & [$-$0.0098, $-$0.0065] \\
CoxPH $-$ DeepSurv & $-$0.0027 & 0.0087 & [$-$0.0036, 0.0097] & 0.0095 & 0.0034 & 0.0029 & [0.0018, 0.0040] \\
RSF $-$ DeepSurv & $-$0.0012 & 0.0144 & [$-$0.0027, 0.0158] & 0.0156 & 0.0024 & 0.0111 & [0.0094, 0.0127] \\
\midrule
\multicolumn{8}{l}{\textit{(c) Difference from the best competitor}} \\
\addlinespace[2pt]
Model & & Upper & & $U_j^{\mathrm{IUT}}$ & & & \\
\midrule
CoxPH &  & 0.0087 &  &  &  &  &  \\
RSF &  & 0.0144 &  &  &  &  &  \\
DeepSurv$^\ast$ &  & 0.0027 &  & 0.0034 &  &  &  \\
\bottomrule
\end{tabular}

\end{table}

\begin{table}[tbp]
    \centering
    \footnotesize
    \caption{Results for SFLLD on the test sample ($n=7{,}225$).
        The panels and columns are as in Table~\ref{tab:real-results-unempdur}, with two additional columns in panel~(a).
        Width is Upper minus Lower, and Position is $100(\mathrm{IPCW}-\mathrm{Lower})/\mathrm{Width}$, the IPCW estimate's percentage distance from the lower end.
        Both columns are calculated from the saved results before rounding the displayed endpoints and IPCW estimates.}
    \label{tab:real-results-sflld}
    \setlength{\tabcolsep}{2.5pt}
    \begin{tabular}{@{}lrrcrrrc@{}}
\toprule
\multicolumn{8}{l}{\textit{(a) IBS of each model}} \\
\addlinespace[2pt]
Model & Lower & Upper & 95\% CR & Width & Position (\%) & IPCW & 95\% CI \\
\midrule
CoxPH & 0.152 & 0.153 & [0.149, 0.156] & 0.0013 & 17.1 & 0.152 & [0.149, 0.155] \\
RSF & 0.155 & 0.156 & [0.152, 0.158] & 0.0011 & 14.7 & 0.155 & [0.152, 0.157] \\
DeepSurv & 0.151 & 0.152 & [0.148, 0.155] & 0.0014 & 17.5 & 0.151 & [0.148, 0.153] \\
\midrule
\multicolumn{8}{l}{\textit{(b) IBS difference between two models}} \\
\addlinespace[2pt]
Pair & Lower & Upper & 95\% CR & $U_{jk}$ & $U_{kj}$ & IPCW & 95\% CI \\
\midrule
CoxPH $-$ RSF & $-$0.0027 & $-$0.0023 & [$-$0.0036, $-$0.0014] & $-$0.0016 & 0.0035 & $-$0.0026 & [$-$0.0035, $-$0.0017] \\
CoxPH $-$ DeepSurv & 0.0012 & 0.0015 & [0.0005, 0.0021] & 0.0020 & $-$0.0006 & 0.0013 & [0.0007, 0.0020] \\
RSF $-$ DeepSurv & 0.0036 & 0.0041 & [0.0026, 0.0051] & 0.0050 & $-$0.0027 & 0.0039 & [0.0029, 0.0049] \\
\midrule
\multicolumn{8}{l}{\textit{(c) Difference from the best competitor}} \\
\addlinespace[2pt]
Model & & Upper & & $U_j^{\mathrm{IUT}}$ & & & \\
\midrule
CoxPH &  & 0.0015 &  &  &  &  &  \\
RSF &  & 0.0041 &  &  &  &  &  \\
DeepSurv$^\ast$ &  & $-$0.0012 &  & $-$0.0006 &  &  &  \\
\bottomrule
\end{tabular}

\end{table}

The IUT at the 5\% significance level certified the prespecified model, DeepSurv, as robustly strictly best only in SFLLD (panel~(c) of Tables~\ref{tab:real-results-unempdur}--\ref{tab:real-results-sflld}).
In all three data sets, DeepSurv had the smallest IPCW estimate on the validation sample and was therefore prespecified as the evaluated model.
For SFLLD, the estimated upper bound on its difference from the best competitor was $-0.0012$, and the one-sided upper confidence bound $U_j^{\mathrm{IUT}}=-0.0006$ remained below zero.
For UnempDur and KKBOX, the IUT did not reject because the estimated upper bounds themselves were positive, 0.0161 and 0.0027, before any allowance for sampling uncertainty.
These nonrejections do not establish that a competitor has a smaller population IBS than DeepSurv.

\section{Conclusion}\label{sec:conclusion}

This paper has derived sharp bounds on the IBS of survival prediction models and on IBS differences between them without restricting the dependence between event and censoring times.
Such bounds are needed because the independent censoring assumed by IPCW cannot generally be verified from right-censored data.
Using the fact that each individual's event time is shared across evaluation times and models, the bounds cover a single model, a pair of models, and, through its upper bound, the difference from the best competitor (Propositions~\ref{prop:ibs-bounds}, \ref{prop:pair-bounds}, and~\ref{prop:best-bounds}).
For a fixed observed-data distribution, we constructed estimators of these bounds, confidence regions, a test of robust strict dominance, and an intersection--union test.
With these tools, a model can be shown to be better under every full-data distribution compatible with the observed data, without specifying the distribution of the unobserved event times.

The simulations and the applications showed that the bounds separate rankings supported by the observed data from rankings that rely on independent censoring.
In the simulations, the confidence regions and tests performed close to their nominal levels, whereas the IPCW intervals often missed the full-data IBS under dependent censoring, although the IUT was far more conservative when two competitors were equally close.
In the applications, the tests at the 5\% level, without adjustment for multiple testing, supported the IPCW ranking and certified DeepSurv as robustly strictly best only for SFLLD, in which 0.4\% of the individuals were censored before the horizon.
For UnempDur and KKBOX, with 56.3\% and 22.4\% censored, no test rejected, so their IPCW rankings were not supported without assuming independent censoring.
Estimated bounds should therefore be reported alongside IPCW estimates.

The method has limitations that point to directions for future work.
Its inference is asymptotic for a fixed observed-data distribution, so its level need not hold uniformly near distributions at which an upper bound is close to zero or several competitors nearly attain the maximum.
It also requires models trained separately from the evaluation sample and comparisons specified in advance (Assumptions~\ref{assump:independent-training} and~\ref{assump:prespecified}).
Moreover, the identified sets can be wide when many individuals are censored before the horizon.
Future work includes narrowing them with weak assumptions credible in the application, such as administrative censoring times known for all individuals.
It also includes extending the bounds to measures other than the IBS, such as the concordance index \parencite{1982Harrell_EvaluatingYieldMedical,2011Uno_CstatisticsEvaluatingOverall}.

\printbibliography[title={References},heading=bibintoc]

\ifproofsinappendix
  \clearpage
  \appendix
  \section*{Appendix}
  \section{Proofs for Section~\ref{sec:partial-identification}}\label{app:proofs-partial-identification}

\IfFileExists{\jobname-pratendpartial-identification.tex}{\printProofs[partial-identification]}{}

  \section{Proofs for Section~\ref{sec:estimation-inference}}\label{app:proofs-inference}

\IfFileExists{\jobname-pratendinference.tex}{\printProofs[inference]}{}

\fi

\end{document}